\documentclass[12pt,preprint]{aastex}
\usepackage{amssymb}

\begin{document}
\title{Mid Infrared Properties of Low Metallicity Blue Compact Dwarf Galaxies from {\em Spitzer/IRS}}

\author{
Yanling~Wu,\altaffilmark{1},
V.~Charmandaris\altaffilmark{2,1,3},
Lei~Hao\altaffilmark{1},
B.~R.~Brandl\altaffilmark{4},
J.~Bernard-Salas\altaffilmark{1},
H.~W.~W.~Spoon\altaffilmark{1},
J.R.~Houck\altaffilmark{1}
\email{wyl@astro.cornell.edu, vassilis@physics.uoc.gr,
  haol@isc.astro.cornell.edu, brandl@strw.leidenuniv.nl,
  jbs@isc.astro.cornell.edu, spoon@isc.astro.cornell.edu, jrh13@cornell.edu}
}

\altaffiltext{1}{Astronomy Department, Cornell University, Ithaca, NY  14853, USA}
\altaffiltext{2}{University of Crete, Department of Physics, P.~O. Box
  2208, GR-71003, Heraklion, Greece}
\altaffiltext{3}{Chercheur Associ\'e, Observatoire de Paris, F-75014,
  Paris, France}

\altaffiltext{4}{Leiden Observatory, 2300 RA Leiden, The Netherlands}

\begin{abstract}
  We present a {\em Spitzer}-based mid-infrared study of a large
  sample of Blue Compact Dwarf galaxies (BCD) using the Infrared
  Spectrograph (IRS), including the first mid-IR spectrum of IZw18,
  the archetype for the BCD class and among the most metal poor
  galaxies known.  We show the spectra of Polycyclic Aromatic
  Hydrocarbon (PAH) emission in low-metallicity environment. We find
  that the equivalent widths (EW) of PAHs at 6.2, 7.7, 8.6 and
  11.2\,$\mu$m are generally weaker in BCDs than in typical starburst
  galaxies and that the fine structure line ratio, [NeIII]/[NeII], has
  a weak anti-correlation with the PAH EW. A much stronger
  anti-correlation is shown between the PAH EW and the product of the
  [NeIII]/[NeII] ratio and the UV luminosity density divided by the
  metallicity.  We conclude that PAH EW in metal-poor high-excitation
  environments is determined by a combination of PAH formation and
  destruction effects.

\end{abstract}

\keywords{dust, extinction -- galaxies: abundances -- galaxies: starburst -- galaxies: 
  dwarf -- infrared: galaxies}

\section{Introduction}

Galaxies formed in the early Universe are likely to have very
different properties from typical evolved galaxies in the local
Universe because they were formed from an environment deficient in
heavy elements.  Important as they are, those primeval galaxies are
beyond the capability of current mid-IR instruments, thus motivating
an ongoing effort to identify local analogs of the galaxy formation
process in the early Universe.

Blue compact dwarfs are a group of extra-galactic objects with faint,
blue optical colors, small sizes and low metallicities. They are
generally dominated by one or more recent bursts of star
formation. Early studies hinted that BCDs have dramatically different
properties compared to normal dwarf galaxies
\citep{Arp1965,Zwicky1966}. Accumulated observational evidence over
the recent years provided more details on the unique properties of
these galaxies \citep [for a review see][]{Kunth2000}.  Early ground
based observations by \citet{Roche1991} on the mid-IR spectra of 60
galaxies revealed that PAH emission is generally suppressed in
low-metallicity galaxies, which could be due to hard photons
destroying the particles that produce the unidentified infrared
bands. \citet{Thuan1999} has shown that there is no sign of these
bands in the mid-IR spectrum of SBS0335-052E taken by the Infrared
Space Observatory (ISO). The suppression of the PAH emission is also
seen in the mid-IR spectra of 4 BCDs discussed by
\citet{Madden2000,Galliano2005,Madden2005}. \citet{Dwek2004} proposed
that the delayed injection of carbon molecules into the interstellar
medium(ISM) might be partly responsible for the absence of PAH
features in young star forming regions, or for the existence of a
metallicity threshold below which PAHs have not formed.
\citet{Hogg2005} suggested that the lack of PAH emission is closely
related to the low luminosity of their sample. More recent work based
on {\em Spitzer} observations has confirmed that PAH emission is
missing in the most metal-poor galaxies \citep{Houck2004b,
Engelbracht2005}.  \citet{Hunt2005} studied the global spectral energy
distributions (SEDs) of 7 BCDs and confirmed that those SEDs deviate
significantly from the standard templates of ``classical'', evolved
and massive starburst galaxies, in the sense that the far-infrared
(FIR) spectra peak at/or shortward of 60\,$\mu$m and mid-IR spectra
show little or no PAH emission. Having a relatively low metallicity,
these galaxies are at an early epoch of their evolution, making them
similar to samples of the distant, more massive protogalaxies, thus
allowing us to study the star formation and chemical enrichment in an
environment likely to be similar to that in the early Universe.

Recently, mid-IR imaging and spectroscopy of a handful of such systems
revealed the presence of dusty embedded sources as well as fine
structure line emission in their spectra, consistent with a hard
radiation field from massive young stars
\citep{Thuan1997,Madden2000,Madden2005}. In general, short bursts (1-5
Myr) of intensive star formation are found to best match the data
\citep{MasHesse1999}. Some BCDs appear to be very rapid star formers,
and thus true dwarf analogs of giant starbursts \citep{Kunth2000}. The
profiles and strengths of the PAHs, which might be good tracers
of star formation, vary substantially \citep{Forster2004}.

Using the unprecedented sensitivity of the Infrared Spectrograph
\citep{Houck2004a} on the {\em Spitzer} Space Telescope
\citep{Werner2004}, we compiled a large sample of of the lowest
metallicity galaxies known. Since little information on their mid-IR
fluxes were available from literature, a large fraction of them were
first observed with the peak-up cameras at 16 and 22\,$\mu$m. As we
discuss in detail in the following section, those which were bright
enough were observed spectroscopically.  In this paper we present
5.2--36\,$\mu$m spectra of 12 galaxies for which the signal-to-noise
ratio (SNR) was high enough to perform a quantitative analysis on the
strength of the spectral features, as well as 16 and 22\,$\mu$m broad
band imaging for the remaining of the sample, covering metallicities
from 1/50 Z$_{\odot}$ to 0.65 Z$_{\odot}$. Many of the remainders are
scheduled for re-observation to achieve higher SNR.  In Section 2, we
describe the sample, the observing strategy, and the data
reduction. Section 3 presents the spectral features observed in each
source while our analysis on these properties is shown in Section
4. We summarize our conclusions in Section 5.  Throughout this work,
we assume a $\Lambda$CDM cosmology with
H$_0$\,=\,71~km~s$^{-1}$\,Mpc$^{-1}$, $\Omega_m$\,=\,0.3 and
$\Omega_{\lambda}\,=\,0.7$.

\section{Observations and Data Reduction}

A total of 61 BCDs have been observed as part of the {\em IRS}
Guaranteed Time Observation (GTO) program. The sample included targets
from the Second Byurakan Survey (SBS), Bootes Void galaxies
\citep{Kirshner1981, Popescu2000}, and several well-known BCDs. Some
basic properties of these galaxies and the observing parameters, such
as: source names, right ascension (RA), declination (Dec), redshift,
{\em Spitzer} aorkeys, observation date, and on-source integration
time for each module and peak-up mode are listed in Table \ref{tab1}.

We used all four {\em IRS} modules, Short-Low (SL, 5.2-14.5\,$\mu$m),
Long-low (LL, 14.0-38.0\,$\mu$m), Short-High (SH, 9.9-19.6\,$\mu$m)
and Long-high (LH, 18.7-37.2\,$\mu$m) to obtain spectra for 26 sources
which were expected to be sufficiently bright for
spectroscopy. SBS1200+589B was the only one of the targets that was
not observed in SL.  The low resolution modules (SL, LL) produce
spectra with a resolution of 64--128 while the high-resolution modules
(SH, LH) have a resolution of $\sim$600
\citep[see][]{Houck2004a}. Nearly all targets were acquired using red
peak-up (RPU) mode. In the process, an image of the source at 22$\mu$m
was obtained in Double Correlated Sampling (DCS) mode and it was used
to locate the mid-IR centroid of the source which was then offset to
the appropriate slit (see Spitzer Observers Manual for more details).
For BCDs that were too faint for spectroscopy, we only obtained broad
band imaging at 16\,$\mu$m (13.5-18.7\,$\mu$m) and 22\,$\mu$m
(18.5-26.0\,$\mu$m) using both {\em IRS} peak-up cameras in Raw Data
Collection (RAW) mode.  Details on this method of imaging with {\em
IRS}, also called CHEAP for Cornell High-Efficiency Advanced Peak-up,
and its photometric advantages compared to the usual {\em IRS} Peakup
Only mode are discussed in \citet{Charmandaris2004}.  Eight of our
galaxies were so faint in the mid-IR that only an upper limit of
$\sim$0.1mJy could be established for their 16 and 22$\mu$m flux
densities.

The data were processed by the {\em Spitzer} Science Center (SSC) data
reduction pipeline version 11.0 and 11.4\footnote{The usage of data
products from two versions of the pipeline was a result of the
extended time-span over which sample was observed and the delivery of
the files from the SSC.  The difference between the two versions is
very small and it does not introduce any inhomogeneity in the sample
or influence the conclusions of the paper.}. The 2-D image data were
converted to slopes after linearization correction, subtraction of
darks, and cosmic ray removal. The DCS peak-up imaging data were
processed on the ground to remove cosmic rays and the residual noise
of the electronics. Fixed aperture photometry was then performed in
order to obtain the counts of the science target. The conversion to
flux density was based on a number of calibration stars for which
peak-up images, {\em IRS} spectra, and reliable templates are
available \citep{Cohen2003}.  The reduction of the spectral data
started from the intermediate pipeline products (droop files), which
only lacked stray light and flat field correction. Individual
pointings to each nod position of the slit were co-added using median
averaging and for SL and LL spectra, the two apertures were
differenced in order to remove the contribution of the background. The
2-D images were extracted with the Spectral Modeling, Analysis, and
Reduction Tool \citep[SMART Ver. 5.5.1][]{Higdon2004} using a variable
width aperture, which scales the extraction aperture with wavelength
to recover the same fraction of the diffraction limited instrumental
point-spread-function (PSF). The data from SH and LH were extracted
using the full slit extraction method from the median of the combined
images.  Since no sky (off--position) measurements were taken, the
contribution of the sky emission was not subtracted from SH and LH
spectra. Then the spectra were flux calibrated by multiplication with
the Relative Spectral Response Function (RSRF), which was created from
the {\em IRS} standard stars $\alpha$ Lac for SL and LL and $\xi$ Dra
for SH and LH for which accurate templates were available
\citep{Cohen2003}. We built our RSRFs by extracting the spectra of the
calibration stars in the exact same way as the science targets, and
dividing the stellar templates by the extracted stellar spectra. We
produced one RSRF for each nod position in order to avoid systematic
flat field errors.  The signal difference between the nod positions
provide the error estimates.  Finally, the flux calibrated spectra of
each order and module of the low-resolution modules except 1st order
LL (LL1) (20--36$\mu$m) were scaled, using LL1 spectrum to define the
continuum (presented in Fig.1 and 2). The associated photometric
points are also presented on the spectra in Fig.1 and 2.

\section{Results}

From the 26 BCDs observed spectroscopically, only 11 produced spectra
with SNR sufficient for features to be measured (SNR $\geq2$ in the
5--8$\mu$m range of the 2nd order SL, SL2). These spectra are
presented in Fig.1 and 2. We also include the first mid-IR spectrum of
IZw18\footnote{The overall shape of the mid-IR continuum is evident
from Fig. 1, but a measurement of spectral features is challenging due
to the noise in the spectrum.}.  The 22\,$\mu$m flux density of each
target with an aperture radius of 7.2$\arcsec$ measured during its
acquisition using the red peak-up camera (circle), as well as a
``synthetic'' 22\,$\mu$m flux obtained by convolving the spectrum with
the filter response function (square) are indicated in the Fig.1 and
2. The {\em IRAC} 8\,$\mu$m (asterisk) and {\em MIPS} 24\,$\mu$m
(triangle) flux densities are also indicated when available
\citep{Engelbracht2005}. One can see that for all point sources,
including SBS0335-052E, NGC1140, Mrk1499, Mrk1450, CG0598, CG0752,
UM461 and IZw18, the two broad band values for the RPU agree very
well. NGC1140 is marginally saturated at 22\,$\mu$m, but we correct
for this effect based on the PSF profile. The agreement of the spectra
and the photometric measurements indicates the internal consistency of
the calibration of the instrument between spectroscopy and peak-up
photometry when all spectra are scaled to LL1. However, as expected,
for sources which are extended or well above the 340\,mJy saturation
limit of red peak-up in a 8.4 second exposure time, such as NGC1569,
IIZw40, Haro11 and UGC4274, there is a disagreement among these
photometric measurements (See Fig.1 and 2).

It is clearly seen from Fig. 1 and 2 that the forbidden transitions of
[SIV] at 10.51\,$\mu$m and [NeIII] at 15.55\,$\mu$m are visible even
in the low-resolution spectra, while [NeII] at 12.81\,$\mu$m, [SIII]
at 18.71\,$\mu$m and [SiII] at 34.8\,$\mu$m lines are seen in most of
them. The line fluxes used in the analysis are determined from
high-resolution spectra (See Fig.3). The emission from PAHs at 6.2,
7.7, 8.6 11.2 and 12.8\,$\mu$m are also clear in several of our
targets and they are indicated in the figures but no pronounced
silicate emission or absorption features are present.  All PAH
measurements presented in this paper were measured from the
low-resolution spectra.

\subsection{Mid infrared Morphology}

As mentioned in Section 2, all spectroscopic observations were
preceded by an acquisition using the 22$\mu$m peak-up camera. This
provided not only more precise position of the objects for telescope
pointing, but also deep 22\,$\mu$m images of our targets.  Analysis of
those images indicates that with the exception of four galaxies:
NGC1569, UGC4274, IIZw40 and Haro11, all other targets are unresolved
at 22\,$\mu$m (FWHM=6.5\arcsec) .  In Fig.4, we display the 22\,$\mu$m
images of these four galaxies, overlayed with the spectrograph slits
as they were placed when the spectra were obtained. Among those
targets NGC1569 shows the most complex structure with two bright cores
in the mid-IR. This source also has very extended emission in the
optical. Therefore, not surprisingly, it requires a large scaling
factor of $\leq$80\% when we ``stitch'' the spectra of 1st order SL
(SL1) and 2nd order LL (LL2), mostly due to the different slit widths
and orientations between the two low-resolution modules. In UGC4274,
we find that most of its 22\,$\mu$m flux originates from a compact
unresolved knot while a similar peak is seen in the near infrared
which is offset by $\sim3\arcsec$ from the mid-IR one.

\subsection{Individual Objects}

{\bf IZw18} --- This is the first identified member of the blue
compact dwarf galaxy class \citep{Searle1972} and until
recently\footnote{Only this year (2005) measurements of
\citet{Izotov2005} showed that the brightest star formation region of
the western component of SBS0335-052 has an oxygen abundance of only
12+log(O/H)=7.12$\pm$0.03 ($\sim$0.017 Z$_{\odot}$), suggesting it is
the most metal poor galaxy known.} held the record of a galaxy with
the lowest metallicity at Z $\sim$0.02 Z$_{\odot}$
\citep{Skillman1993}. IZw18 is located at a distance of 12-15 Mpc and
is a bona fide young galaxy and no red giant branch stars are
seen\citep{Izotov2004}. \citet{Zwicky1966} described it as a double
system of compact galaxies, which are in fact two star-forming
regions, a northwest component and a southeast one, separated by an
angular distance of 5.8\arcsec. Examination of the spatial
distribution of the stellar populations suggest that the star
formation process is still building the main body from inside out
\citep{Izotov2004}. Both \citet{Izotov1997a} and \citet{Legrand1997}
detected several broad emission components in this BCD, suggesting
that Wolf-Rayet (WR) features can also exist in extremely low
metallicity environment. It is intrinsically very faint in the
infrared. The first mid-IR spectrum of IZw18 is presented in Fig.1. As
we discuss in the subsequent sections we are able to estimate the
mid-IR spectral slope from its low-resolution spectrum. The fine
structure lines of [SIV] and [NeIII] can clearly be seen in its
high-resolution spectrum, though the identification of [NeII] and
[SIII] is not yet firm. A much deeper observation with more exposure
time has been scheduled and the new spectrum will be presented in a
future paper \citep{Wu2006}.\\

{\bf SBS0335-052E} --- Currently the third most metal-poor galaxy
known with 0.024 Z$_{\odot}$ \citep{Izotov1997b}, SBS0335-052E is at a
distance of 58 Mpc. \citet{Thuan1997} found that stars in SBS0335-052E
are younger than $\sim$100 Myr and the current burst of star formation
is no older than 5 Myrs, thus making it a truly young system. This
galaxy is unexpectedly bright in the mid-IR
\citep{Thuan1999,Houck2004b} with roughly 75\% of the total luminosity
coming from the mid-IR \citep{Plante2002}. It has six compact regions
of massive star formation, five of which are visible and one obscured,
and all of them lying within a diameter of 526 pc \citep{Thuan1997}.
There is no sign of PAH emission in its spectrum and silicate
absorption at 9.7\,$\mu$m is clearly seen. The SED of the galaxy is
dominated by a very strong continuum, which unlike typical star
forming galaxies, such as NGC7714 \citep{Brandl2004}, peaks at
$\sim$28 $\mu$m (f$_{\nu}$), indicating the presence of little cold
dust. The fine structure lines of [NeIII] and [SIV] are present,
though not very strong. \\

{\bf UM461} --- This is a dwarf galaxy with a double nucleus and is
found at a distance of 15 Mpc.  It has an external envelope which is
strongly distorted towards the South-West, suggesting a recent tidal
event \citep{Doublier1999}. The metallicity of UM461 is $\sim$0.087
Z$_{\odot}$ \citep{Kniazev2004}. With only 168 seconds of integration
time in SL module, the obtained spectrum is very noisy. However, the
fine structure lines, such as [SIV], [NeIII] and [SIII] can be clearly
seen longward of 10\,$\mu$m. The identification of PAH emission is
doubtful and an upper limit of 0.199\,$\mu$m for the PAH EW at
11.2\,$\mu$m is indicated in Table \ref{tab4}. \\

{\bf Haro11} --- This is a metal poor galaxy with $\sim$0.1
Z$_{\odot}$ \citep{Bergvall2000}. Its distance is $\sim$88 Mpc and it
has an infrared luminosity, L$_{\rm IR}$ of 1.8$\times$10$^{11}$
L$_{\odot}$\footnote{Calculated from the {\em IRAS} flux densities
following the prescription of \citet{Sanders1996}: L$_{\rm
IR}$=5.6$\times$10$^5$D$_{Mpc}^2$($13.48S_{12}+5.16S_{25}+2.58S_{60}+S_{100}$).}.
Haro11 is a moderately strong radio source, with spatially extended
continuum emission at 6 and 20
cm\citep{Vader1993,Heisler1995}. Multiple nuclei are apparent on
optical broad band images \citep{Heisler1994} as well as narrow-band
H$\alpha$ images \citep{Heisler1995}. All three nuclei are similar in
continuum stellar emission \citep {Heisler1995} and spectra of the
nuclei confirm that they are all at the same distance
\citep{Vader1993}.  The PAH features are much weaker compared with
NGC1140 but the fine structure lines of [SIV], [SIII], [NeIII] and
[NeII] lines are clearly present. \\

{\bf Mrk1450} --- A dwarf compact object with a projected dimension
less than 1 kpc, Mrk1450 is located at a distance of $\sim$14 Mpc. Its
optical images display moderately deformed circular isophotes and a
central star forming component. A strong color gradient is present
reaching a $B-R\sim$2.2 mags at the outskirts of the galaxy, while the
surface brightness distribution obeys an r$^{1/4}$ law
\citep{Doublier1997}.  This galaxy has an L$_{\rm IR}$\footnote{For
this galaxy, the {\em IRAS} 12\,$\mu$m and 25\,$\mu$m flux densities
are calculated by convolving our spectrum with the {\em IRAS}
filters. The 100\,$\mu$m flux density is an upper limit, thus making
the L$_{\rm IR}$ also an upper limit.} less than 1.8$\times$10$^8$
L$_{\odot}$. Its metallicity is $\sim$0.12 Z$_{\odot}$
\citep{Izotov1999}. Due to the low SNR in SL2, PAH emission can not be
identified clearly below 10$\mu$m, even though the 11.2\,$\mu$m and
possibly 12.8\,$\mu$m PAHs are present. The fine structure lines of
[SIV] and [NeIII] are strong. \\

{\bf IIZw40} --- This is a prototypical \ion{H}{2} galaxy
\citep{Sargent1970} with L$_{\rm IR}$\footnote{For this galaxy, only
upper limit exists for {\em IRAS} 100\,$\mu$m, thus making the L$_{\rm
IR}$ also an upper limit.}  less than 2.9$\times$10$^9$ L$_{\odot}$ at
a distance of $\sim$10~Mpc.  The metallicity of this galaxy is
$\sim$0.17 Z$_{\odot}$ \citep{Cervino1994}. It consists of a compact,
extremely bright core and two fan-like structures, which have been
interpreted as the result of a merger between two small galaxies
\citep{Baldwin1982,Brinks1988}. Star formation is concentrated in the
nucleus while the double structure is quite red and shows no star
formation activity \citep{Cairos2001}. H$\alpha$ emission is very
strong, contributing 40\% to the $R-$band flux at the nucleus. The
starburst could be very young and its strength is rather extraordinary
\citep{Deeg1997}. It also displays WR features \citep{Conti1991}. Our
mid-IR spectrum indicates that the PAHs are present, though extremely
weak. The fine structure lines of [SIV], [SIII], [NeII] and [NeIII],
have been detected.\\

{\bf NGC1569} --- A nearby dwarf galaxy \citep[D$\sim$2.2 Mpc
][]{Israel1988} which has a metallicity of $\sim$0.19 Z$_{\odot}$
\citep{Kobulnicky1997} and is currently in the aftermath of a massive
burst of star formation \citep{Waller1991}, with L$_{\rm IR}$ of
5.8$\times$10$^8$ L$_{\odot}$.  It lies close to the plane of the
Galaxy and therefore its optical properties are strongly affected by
the Galactic dust extinction \citep{Kinney1993}. The presence of a
broad emission feature around 4650 {\AA} and a broad base to the
H$\alpha$ line indicate that WR stars are present in the nucleus of
NGC1569 \citep{Ho1995}. It contains two bright super star clusters.
The very hot, bright nucleus of this galaxy resembles a superluminous,
young star cluster \citep{Arp1985}. The fine structure lines of [SIV]
and [NeIII] are stronger compared to [SIII] and [NeII], which can be
seen clearly even in the low-resolution spectrum (See Fig.2). The PAH
features are clearly present.  As discussed by
\citet{Galliano2003,Galliano2005}, the mid-IR emission is dominated by
small grains.\\

{\bf Mrk1499} --- This is an irregular galaxy displaying a very blue
elongated central structure with two components in the optical, the
brightest being off-centered with respect to the outer contours
\citep{Doublier1997}. These authors suggested that this structure is
reminiscent of ``double nuclei'' objects, while \citet{Petrosian2002}
argued that this could be just an observational artifact.  It has an
L$_{\rm IR}$\footnote{For this galaxy, only {\em IRAS} 60\,$\mu$m and
100\,$\mu$m are available. For {\em IRAS} 12\,$\mu$m and 25\,$\mu$m,
we convolved our spectrum with the two {\em IRAS} filters and
calculated their values.} of 1.3$\times$10$^9$ L$_{\odot}$ and is at a
distance of $\sim$38 Mpc. An oxygen abundance of $\sim$ 0.3
Z$_{\odot}$ has been derived from the line measurements of
\cite{Petrosian2002}, using the N2 calibrator \citep{Denicolo2002},
which allows us to calculate the oxygen abundance based on the
[NII]/H$\alpha$ ratio.  PAH features are prominent and several fine
structure lines are detected. In this particular galaxy, the SNR in
SL2 is rather poor, thus making the definition of the continuum
challenging. We measured the 6.2\,$\mu$m PAH EW by defining a maximum
and minimum local continuum and then averaging the two
measurements. \\

{\bf NGC1140} ---This is a blue irregular galaxy at a distance of
$\sim$25 Mpc, containing large mass of ionized gas in its center
\citep{Kinney1993}. Its metallicity is $\sim$0.4 Z$_{\odot}$
\citep{Calzetti1997}.  \citet{Lamb1986} concluded that NGC1140 has
experienced a single burst of star formation and its population mainly
consists of main-sequence stars with a contribution from supergiants
and, possibly, WR stars.  \citet{Hunter1994} used the Hubble Space
Telescope ({\em HST}) Planetary Camera to study the central supergiant
\ion{H}{2} region and found that the central 1/2 kpc of NGC1140
contains $\sim$7 blue, luminous, compact super star clusters. It has a
very small average size of grains \citep{Galliano2003,Galliano2005}
and a L$_{\rm IR}$ of 4.3$\times$10$^9$ L$_{\odot}$.  The low
resolution mid-IR spectrum shows pronounced PAH emission, almost
comparable to that in the typical starburst galaxy NGC7714,
while the fine structure lines are stronger compared with NGC7714.\\

{\bf UGC4274} --- A galaxy also known as NGC2537 and Mrk86, UGC4274 is
extended in the mid-IR (see Fig.3) and is located at a distance of
$\sim$7~Mpc. It has a large, very irregular nucleus consisting of
$\sim$80 star-forming knots distributed in a circular region which is
surrounded by a red envelope \citep{GildePaz2000}.  The H$\alpha$
image shows a very complex gas distribution, with multiple filaments,
loops and twisted features \citep{Cairos2001}. It has an L$_{\rm IR}$
of 5.0$\times$10$^8$ L$_{\odot}$ and emission from PAHs dominates its
spectrum. An oxygen abundance of 12+log(O/H)=8.05 (Z$\sim$0.13
Z$_{\odot}$) has been derived by \citet{Meier2001} based on the
relationship between B magnitude and metallicity. Since this method
has a large uncertainty and there is no direct spectroscopically
measured oxygen abundance available, we calculated the metallicity
using the N2 calibrator.  A value of 12+log(O/H)=8.52$\pm0.13$
(Z$\sim$0.41 Z$_{\odot}$) has been derived from the line measurement
in \citet{Ho1997}. All major mid-IR fine structure lines of the galaxy
are very strong.\\

{\bf CG0598} --- This is a galaxy selected from the Bootes Void sample. 
It is at a distance of 253 Mpc and it has an
L$_{\rm IR}$\footnote{The {\em IRAS} 12\,$\mu$m and 25\,$\mu$m
  flux densities are calculated by convolving the spectrum with the
  {\em IRAS} filters. The {\em IRAS} 100\,$\mu$m flux density is an
  upper limit thus making L$_{\rm IR}$ an upper limit as well.} less than
5.9$\times$10$^{10}$ L$_{\odot}$.  In the optical, this galaxy is a
flattened disk system with a bright, round nucleus.  It has a smooth,
elongated surrounding disk which is asymmetric with respect to the
nucleus \citep{Cruzen1997}. The metallicity of this galaxy is $\sim$
0.65 Z$_{\odot}$ \citep{Peimbert1992}. In Fig.2, we can see that
CG0598 shows very strong PAH emission in all bands and the usual fine
structure lines of [NeII], [NeIII], [SIV] and
[SIII] can be identified.\\

{\bf CG0752} --- This galaxy is $\sim$91 Mpc away with an L$_{\rm
  IR}$\footnote{The {\em IRAS} 12\,$\mu$m flux density is calculated
  by convolving the spectrum with the {\em IRAS} filter.} of
  2.2$\times$10$^{10}$ L$_{\odot}$. It has remained uncatalogued for a
  long time because of its proximity to bright star
  \citep{Sanduleak1987}. PAH emission is moderately strong and the
  fine structure lines of [SIV], [SIII], [NeIII] and [NeII] are
  present.  There is no measured oxygen abundance available from
  literature for this galaxy.\\

\section{Analysis}

\subsection{The mid-infrared spectral slope of BCDs}

The photometric properties and metallicities of the 53 galaxies in our
BCD sample (excluding the 8 objects that were too faint in mid-IR) are
listed in Table \ref{tab2}. The 16 and 22\,$\mu$m flux densities of
the sources are based on broad band imaging using the blue or red
peak-up cameras, while the B and K magnitudes are taken from
literature. For the 12 BCDs with good IRS spectra, we also calculated
their ``synthetic'' 16 and 22\,$\mu$m flux densities by convolving the
spectra with the {\em IRS} peak-up filters. For comparison, we list
those values in Table \ref {tab3}. We also present the 22 to
16\,$\mu$m flux density ratio, $f_{22}$/$f_{16}$ ($f_{22} \equiv
f_{\nu}(22\,\mu m$))in Table \ref {tab2}. This ratio provides the
mid-IR slope of the spectrum and can be a useful probe of the shape of
the mid-IR SED even if the galaxy are too faint to obtain a
spectrum. One of our sources, SBS0335-052E, has a particularly low
value of $f_{22}$/$f_{16}$, indicating the absence of cold dust around
the central star forming region. As discussed by \citet{Houck2004b},
the SED of SBS0335-052E peaks maximum f$_{\nu}$ at a much shorter
wavelength of $\sim$ 28\,$\mu$m while most normal galaxies peak
longward of 60\,$\mu$m.

We explore the variation on the $f_{22}$/$f_{16}$ ratio as a function
of metallicity in Fig.5a. In addition to the 12 BCDs, we include in
this figure a typical starburst galaxy NGC7714 \citep{Brandl2004}, and
an Ultra-Luminous Infrared Galaxy (ULIRG), UGC5101 \citep{Armus2004},
in order to sample a wider range of different types of galaxies
forming massive young stars. The BCDs with only photometric CHEAP
images are included using crosses as well when their metallicities are
available from literature.  It has been pointed out by
\citet{Hunter1989} that no clear metallicity dependence is observed in
mid-IR and FIR colors of normal galaxies. This is also confirmed in
Fig.5a where the $f_{22}$/$f_{16}$ ratio for all the BCDs seems to
have an average value of $\sim$2.5 and shows no metallicity
dependence.

In Fig.5b and 5c, we plot the ratio of the flux density at 22\,$\mu$m
to the {\em IRAC} 8\,$\mu$m band, $f_{22}$/$f_{8}$, as well as the
{\em MIPS} 24\,$\mu$m to the {\em IRAC} 8\,$\mu$m, $f_{24}$/$f_{8}$,
as a function of their metallicities. The {\em IRAC} 8\,$\mu$m and
{\em MIPS} 24\,$\mu$m flux densities are calculated by convolving the
mid-IR spectra of these BCDs with the appropriate filter profiles.
\citet{Engelbracht2005} observed that the 8\,$\mu$m to 24\,$\mu$m
color changes markedly between 1/3 to 1/5 solar metallicity and
suggested that this change is predominantly due to a decrease in the
8\,$\mu$m emission. Our Fig.5b and 5c are very similar to Fig.2 in
their paper. It appears that there is a separation on the
$f_{24}$/$f_{8}$ ratio around 1/5 Z$_\odot$ on our plots with the
exception of SBS0335-052, which as mentioned earlier shows no PAH
detection.

It is very important to note that, as shown by our data, using the
$f_{22}$/$f_{8}$, or $f_{24}$/$f_{8}$ ratios to infer the strength of
the PAH emission is not a very robust method. Since the 7.7\,$\mu$m
and 8.6\,$\mu$m PAH emission is included in the {\em IRAC} 8\,$\mu$m
band, one might expect that a lower $f_{22}$/$f_{8}$ ratio would
indicate the presence of a stronger PAH emission from the
galaxy. However, a low $f_{22}$/$f_{8}$ ratio can also be due to
decreased emission at 22\,$\mu$m, resulting from the lack of cooler
dust. A galaxy with a flat continuum but no PAH emission can have the
same $f_{22}$/$f_{8}$ ratio as a galaxy which shows strong PAH
emission and a steep continuum. If we consider SBS0335-052E and
UGC4274, we can see that they have a very similar $f_{22}$/$f_{8}$
ratio, even though no PAHs are detected in SBS0335-052E while PAHs are
prominent in UGC4274. This can be understood when we take into account
the $f_{22}$/$f_{16}$ ratio. It is evident from Fig.5a that these two
sources have very different mid-IR spectral slopes. This explains why
SBS0335-052E lies away from the upper left corner of Fig.5a. We have
also plotted the PAH EW at 6.2\,$\mu$m and 11.2\,$\mu$m as a function
of the $f_{24}$/$f_{8}$ ratio in Fig.6a and 6b. We can see that
overall a lower $f_{24}$/$f_{8}$ value indicates a stronger PAH
emission, but there are some measurements that deviate from this
trend. For completeness we also present the PAH EW at 6.2\,$\mu$m and
11.2\,$\mu$m as a function of $f_{22}$/$f_{16}$ in Fig.6c and 6d but
no clear correlation can be seen.

\subsection{PAH and metallicity}

In Section 1, we mentioned that the absence of PAH emission could be
due to the low abundance of carbon and/or nucleating grains.  In
SBS0335-052E, the most metal poor galaxy for which a high quality
mid-IR spectrum is available, there is an upper limit of just
0.018\,$\mu$m in the EW of the 6.2$\mu$m feature
\citep{Houck2004b}. To examine a possible variation in the PAH EW with
metallicity, we plotted the PAH EW of 6.2\,$\mu$m and 11.2\,$\mu$m for
our sample as function of their metallicity in Fig.7.  The PAH EW were
derived by integrating the flux of the feature in the mean spectra of
both nod positions above an adopted continuum, and then divided by the
average continuum flux in the integration range. The baseline was
determined by fitting a spline function to the selected points.  The
wavelength limits for the integration of the features were
approximately 5.95\,$\mu$m to 6.55\,$\mu$m for the 6.2\,$\mu$m PAH,
7.15\,$\mu$m to 8.20\,$\mu$m for the 7.7\,$\mu$m PAH, 8.20\,$\mu$m to
8.90\,$\mu$m for the 8.6\,$\mu$m PAH and 10.80\,$\mu$m to
11.80\,$\mu$m for the 11.2\,$\mu$m PAH feature. Here we only plot the
6.2\,$\mu$m and 11.2\,$\mu$m PAH EW, but measurements on the
7.7\,$\mu$m and 8.6\,$\mu$m PAH EW can be found in Table
\ref{tab4}. We can see that PAH emission is absent in the most
metal-poor BCDs and its strength is generally suppressed in a low
metallicity environment. This result is contrary to what has been
found in typical starburst galaxies where the PAH EW is much stronger
with values greater than 0.5\,$\mu$m for the 6.2\,$\mu$m and
7.7\,$\mu$m PAH bands \citep[i.e.][]{Brandl2004,Brandl2005}. In Fig.7,
it appears that there is a trend showing that galaxies with a lower
metallicity may have smaller PAH EW. Except for the two galaxies which
have large uncertainty in their oxygen abundance, the PAH EW and
metallicity seem to correlate quite well. The errors in the EW
indicated in Fig.7 vary from one source to the other and are mainly
due to the different SNR among the sources.  Quantifying this relation
by a least-squares (logarithmic) fit gives us a slope of
N=1.57$\pm0.63$ for the 6.2\,$\mu$m PAH EW and N=1.16$\pm0.36$ for the
11.2\,$\mu$m PAH EW. This weak trend we are seeing that the PAH EW is
probably related to metallicity effects agrees with the relation
suggested by \citet{Hogg2005} and \citet{Engelbracht2005}.

\subsection{The hardness of the radiation field}

The presence of a young starburst in a low metallicity environment
results in the production of high energy photons which can propagate
relatively large distances before being absorbed by the metals in the
ISM.  High-resolution {\em IRS} spectra were taken of 26 galaxies and
the results for the 12 BCDs that we discussed above are summarized in
Fig.3 and Table \ref{tab4}. Because of the rather large difference in
the ionization potentials of Ne$^{++}$ (41eV) and Ne$^+$ (22eV), the
ratio of [NeIII]/[NeII] is a good tracer of the hardness of the
interstellar radiation field \citep{Thornley2000}. The [SIV]/[SIII]
ratio, another indicator of the hardness of the radiation field has
also been studied in this paper. Given the fact that extinction in
mid-IR is only a few percent of the optical extinction
\citep{Draine2003}, those line ratios are far less sensitive to the
differential extinction compared to optical or UV lines. We plot the
two line ratios in Fig.8.  Since the extinction effects on [NeIII]
(15.55\,$\mu$m) and [NeII] (12.81\,$\mu$m) are similar, the ratio of
[NeIII]/[NeII] is only weakly affected by extinction, while for
[SIV]/[SIII], we indicate the effect of differential extinction by the
arrow on the upper left corner of the plot for an A$_v$ of $\sim$13
mag.

We also studied the dependence of [NeIII]/[NeII] ratio on metallicity
and the results are presented in Fig.9. We find that our 11 BCDs (CG0752, 
which has no metallicity measurements in literature, is not included.)
appear to define a band, where galaxies with lower metallicity display
a higher [NeIII]/[NeII] ratio, thus harboring a harder radiation
field.  For a given [NeIII]/[NeII] ratio,  there is significant 
scatter within metallicities.  This scatter could be due to local variations 
or differential/patchy dust extinction in these systems. It also suggests
that the metallicity may not be the only factor that affects the
hardness of the radiation field. The results on the [SIV]/[SIII] ratio are
plotted in Fig.10 and we can see that it shares a similar trend as the
[NeIII]/[NeII] ratio. For the discussion that follows we choose the
[NeIII]/[NeII] ratio since it is less dependent on the extinction and
the neon lines are generally stronger. Further more, the [NeIII] and
[NeII] lines span a wider range in ionization potential than [SIV] and
[SIII].

\subsection{PAH and the [NeIII]/[NeII] ratio}

As discussed by \citet{Madden2000}, PAHs are at best a minor dust component 
in dwarf galaxies. In Fig.11, we plot the EW of the 6.2 and 11.2\,$\mu$m PAHs 
as a function of the hardness of the radiation field, using [NeIII]/[NeII]
as an indicator. We observe that the PAH EWs at 6.2 and 11.2\,$\mu$m 
are generally suppressed in a harder radiation field as indicated by
a larger [NeIII]/[NeII] ratios, suggesting that the deficiency
in PAH emission may be related to the destruction of the
photodissociation region (PDR) by hard UV photons. However, the
trend we see is weak. A conventional least-squares fit that minimizes 
(logarithmic) residuals in PAH EW yields a slope of N=-0.91$\pm0.31$ 
for the 6.2\,$\mu$m PAH and N=-0.92$\pm0.27$ for the 11.2\,$\mu$m PAH.  
In addition, we found that for a given [NeIII]/[NeII] ratio, we can have 
different PAH EW values. For example Haro11 has a considerably weaker 
PAH EW at both 6.2\,$\mu$m and 11.2\,$\mu$m as compared with NGC1140 even 
though they have a very similar [NeIII]/[NeII] ratio. This scatter 
indicates that some other or additional process is in play. 

\subsection{PAH and luminosity density}

Another important parameter of the radiation field is its UV
luminosity density. Does this also play a role in the destruction of
PAH molecules? In young starbursts, L$_{FIR}$ is representative of the
total UV luminosity. We estimate the luminosity density of the 12 BCDs
in our sample, using the 22\,$\mu$m luminosity.  \citet{Takeuchi2005}
show that there is linear relation between the mid-IR luminosities
(12\,$\mu$m or 25\,$\mu$m) and the total infrared luminosity, even
though \citet{Dale2005} has shown a peak-to-peak uncertainty of a
factor of 5 in using the 24\,$\mu$m luminosity to indicate the total
infrared luminosity for their sample of starburst galaxies.  We
calculate L$_{22\mu m}$ using the 22$\mu$m images obtained with the
red peak-up camera and divide it by the estimated volume (V) of the
objects from their optical or infrared images\footnote{We use the
mid-infrared images to estimate the volume for sources that are
resolved at 22\,$\mu$m and the optical images for the rest of the
sample.}. This should be proportional to the UV luminosity density.
The results are listed in Table \ref{tab5}.  We present the 6.2 and
11.2$\mu$m PAH EW as a function of the 22$\mu$m luminosity density in
Fig.12. There are fewer data points on the plot compared to the table
since we do not have PAH EW measurement for all the galaxies in Table
\ref{tab5}. We can see that there is a trend that PAH EW decreases
with increasing luminosity density. However, this relation is somewhat
weak. A least-squares fit (logarithmic) of the slope returns
N=-0.63$\pm0.21$ for the 6.2\,$\mu$m PAH and N=-0.77$\pm0.18$ for the
11.2\,$\mu$m PAH.  There is considerable scatter on this plot. For
instance, SBS0335-052E and Haro11 have very similar $L_{22\mu m}$/V
value, but we have not found any PAH emission in SBS0335-052E while
PAHs are weak but clearly present in Haro11.

A plot of the dependence of the PAH EW as a function of the product of
the hardness of the radiation field and the luminosity density,
[NeIII]/[NeII]*L$_{22\mu m}$/V is presented in Fig. 13. We observed
that it is better correlated than both the PAH EW vs [NeIII]/[NeII]
and PAH EW vs luminosity density.  A least-squares fit (logarithmic)
returns N=-0.39$\pm0.12$ for 6.2\,$\mu$m PAH and N=-0.47$\pm0.10$ for
11.2\,$\mu$m PAH. This suggests that an increased density of harder
photons destroys more PAH molecules. The fact that we are seeing a
better correlation on Fig.13 than in Fig.11 or Fig.12 suggests that
both the luminosity density and the hardness of the radiation field
contribute to the destruction of PAHs.

\subsection{PAHs: formation and destruction effects}

We have shown in the previous section that there are generally weaker
PAHs in metal-poor environments. We have also discussed the effects of
the hardness of the radiation field and the luminosity density on the
destruction of PAH molecules. Is the absence of PAHs in metal-poor
galaxies solely due to formation effects or destruction effects or
some combination? To examine this we plot the PAH EW as a function a
new quantity: ([NeIII]/[NeII])$\times$(L$_{22\,\mu
m}$/V)$\times$(1/Z), where the product of the neon ratio and the
luminosity density represents the destruction effect and Z, the
metallicity of the galaxy, represents the formation effect. We can see
in Fig. 14 that there is a much tighter anti-correlation on this plot
as compared to the previous series of plots on the PAH EW we have
shown.  A least-squares fit (logarithmic) gives a slope of
N=-0.39$\pm0.09$ for the 6.2\,$\mu$m PAH and N=-0.43$\pm0.08$ for the
11.2\,$\mu$m PAH. We conclude that both the formation effects (Z) and
the destruction effects ([NeIII]/[NeII])$\times$(L$_{22\,\mu m}$/V)
contribute to the weak PAH EW.

\section{Conclusions}

We have explored the mid-IR properties of blue compact dwarf galaxies
with {\em Spitzer IRS}. We obtained broad band images at 16 and
22\,$\mu$m from the {\em IRS} peak-up camera. Using the low-resolution
{\em IRS} spectra, we detected emission from PAHs, at 6.2, 7.7, 8.6,
11.2 and 12.8\,$\mu$m, for most of our galaxies though the strength
varies considerably.  We also detected a number of fine structure
lines in our mid-IR spectra, including [SIV], [NeII], [NeIII] and
[SIII] and found that for metallicities ranging between 1/50
Z$_{\odot}$ to 0.65 Z$_{\odot}$, the line ratios, [NeIII]/[NeII] and
[SIV]/[SIII], which measure the hardness of the ionization field, vary
inversely with metallicity. Our study yielded the following
conclusions:

1) The ratio of $f_{22}$/$f_{16}$ is $\sim$2.5$\pm0.6$ on average, but
it does not show any dependence on metallicity.

2) Both the emission of PAHs and a hot dust component affect the
$f_{22}$/$f_{8}$ (or $f_{24}$/$f_{8}$) ratio in a similar way. These
effects are difficult to disentangle without mid-IR spectra or
measurements of the mid-IR fluxes at several wavelengths. Predictions
of the strength of PAHs based on a single pair of mid-IR broad band
fluxes may be problematic and should be used with caution (see Fig. 5
and 6).

3) The emission of PAHs in metal-poor BCDs is generally suppressed
   (see Fig 7).

4) The product of the hardness of the radiation field as traced by the
neon line ratio with the mid-IR luminosity density, a ``stand in'' for
the intensity of UV luminosity, correlates well with the PAH EW (see
Fig. 13).

5) The product of the hardness of the radiation field and the
luminosity density divided by the metallicity of the galaxies has a
strong correlation with the PAH EW. This suggests that the absence of
PAHs in metal-poor environment is due to a combination of formation
effects and destruction effects (Fig.14).

\acknowledgements 

We would like to thank D. Devost, L. Armus, Aigen Li and G.S. Sloan
for insightful discussions. We also thank the anonymous referee whose
careful reading and detailed comments greatly improved this
manuscript. The IRS was a collaborative venture between Cornell
University and Ball Aerospace Corporation funded by NASA through the
Jet Propulsion Laboratory and the Ames Research Center.  Support for
this work was provided by NASA through Contract Number 1257184 issued
by JPL/Caltech.

\clearpage

\begin{figure*}
  \epsscale{1.0}
  \plotone{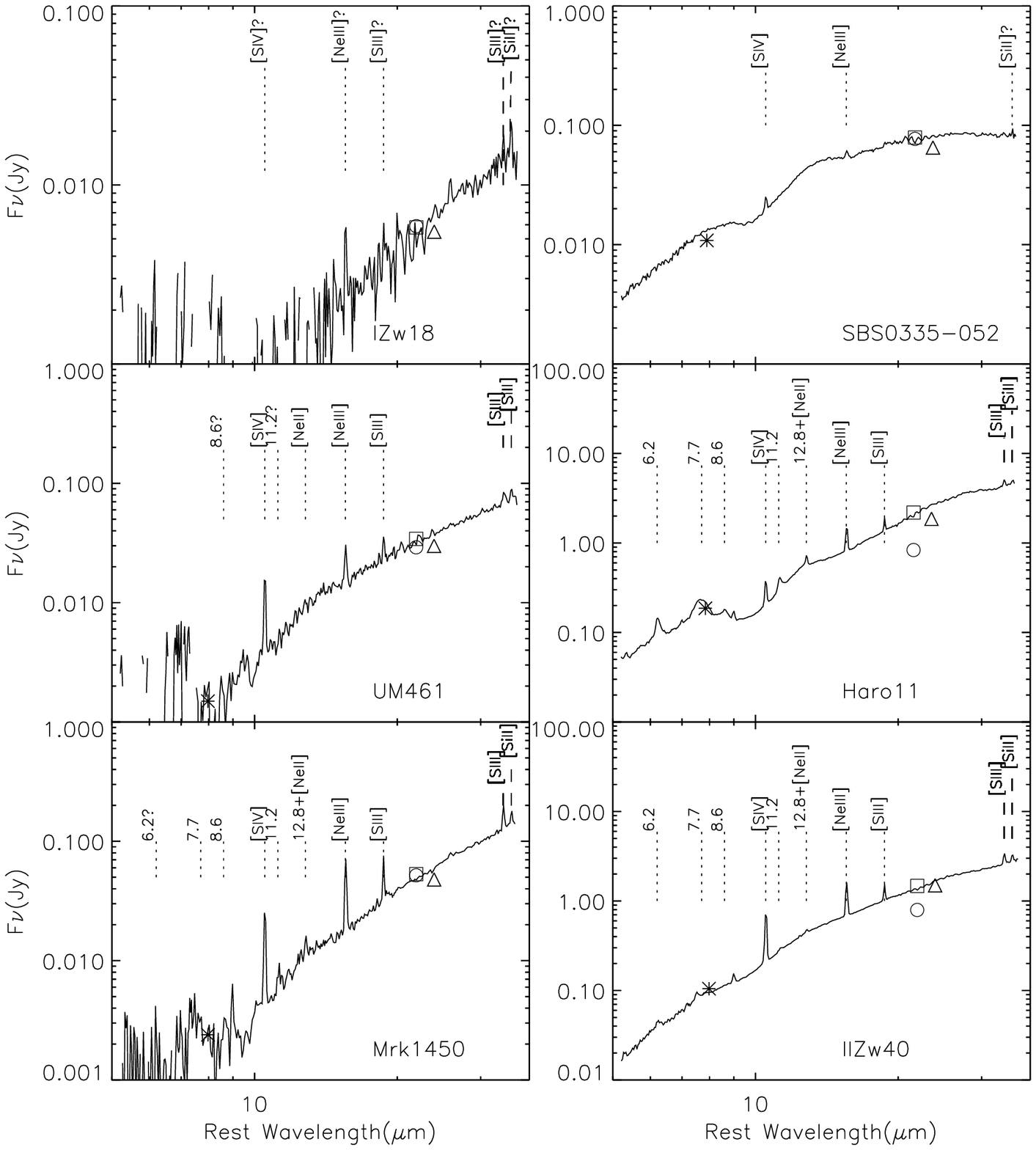}
  \caption{{\em Spitzer}/IRS mid-IR spectra of six BCDs: IZw18,
    SBS0335-052E, UM461, Haro11, Mrk1450 and IIZw40. The spectra of
    different orders were stitched and scaled to match the flux
    density in the 1st order LL. The circles represent the photometric
    flux from 22$\mu$m red peak-up camera, which has an uncertainty of
    6\%.  The squares indicate the flux density at 22\,$\mu$m measured
    from the spectra using ``synthetic'' method (See the text for
    detail). The asterisk and the triangle indicate the {\em IRAC}
    8\,$\mu$m and {\em MIPS} 24\,$\mu$m measurements respectively.}
  \label{fig:fig1}
\end{figure*}

\begin{figure*}
  \epsscale{1.0}
  \plotone{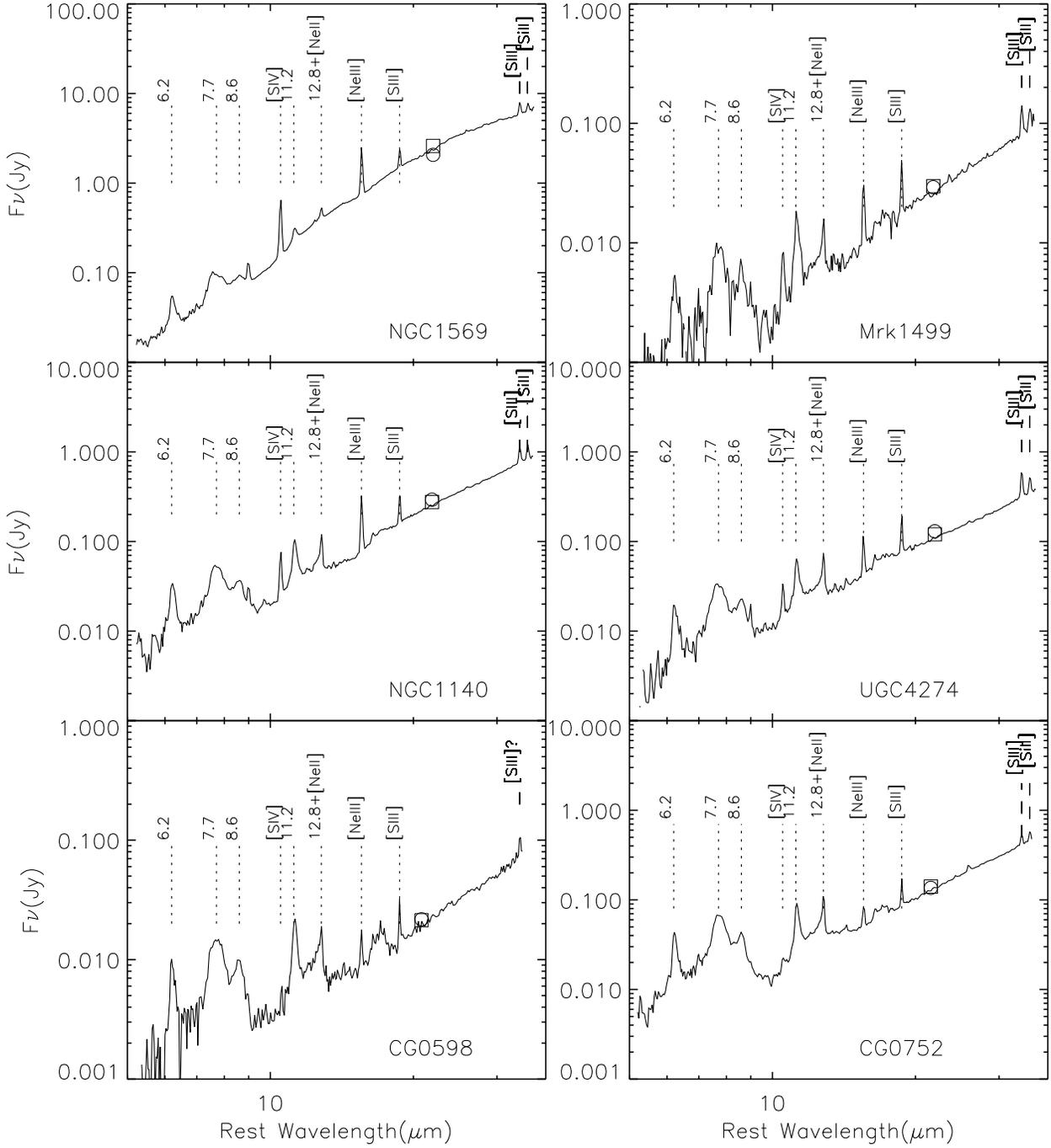}
  \caption{Same as Fig.1 but for another six BCDs: NGC1569, Mrk1499,
    NGC1140, UGC4274, CG0598 and CG0752.}
  \label{fig:fig2}
\end{figure*}

\begin{figure}
  \epsscale{1.0}
  \plotone{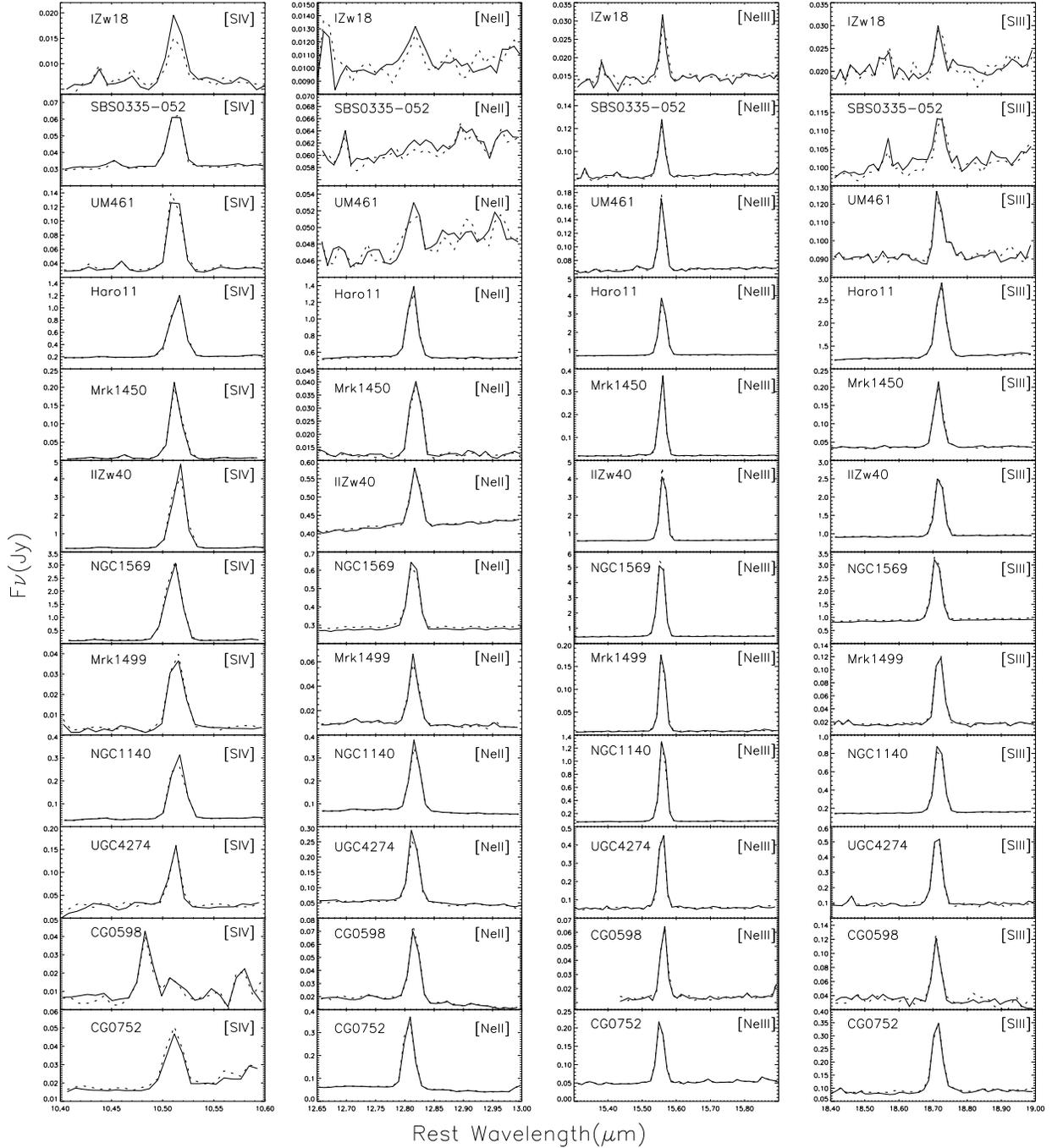}
  \caption{The mid-IR fine structure lines of [SIV] (10.51\,$\mu$m),
    [NeII] (12.81\,$\mu$m), [NeIII] (15.55\,$\mu$m) and [SIII]
    (18.71\,$\mu$m) from the high resolution spectra of the twelve
    BCDs. The solid and dotted lines denote the spectra from the first
    and second nod positions of the slits. Note that sky emission has
    not been subtracted and no stitching and scaling between different
    modules have been done.}
  \label{fig:fig3}
\end{figure}

\begin{figure*}
  \epsscale{1.0}
  \plotone{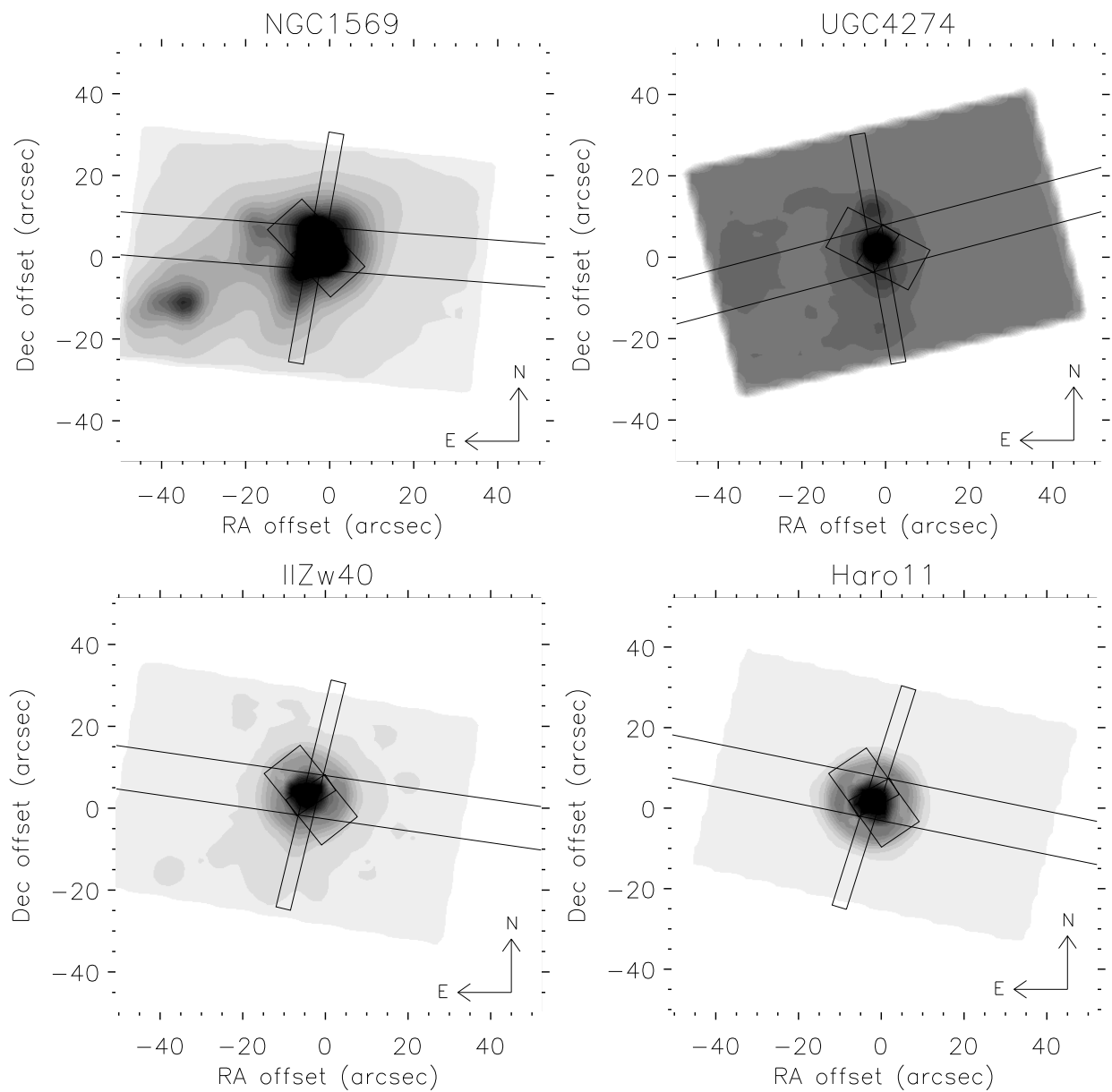}
  \caption{The 22\,$\mu$m peak-up images of the four sources from our
    sample, NGC1569, UGC4274, IIZw40 and Haro11, which display
    extended emission. The images are overlayed with the {\em IRS}
    slits at the location from where the spectrum was obtained. The
    widths of the slits increase in the order of: SL, SH, LL, LH and
    their length in the order of: SH, LH, SL, LL. }
   \label{fig:fig4}
\end{figure*}

\begin{figure}
  \epsscale{1.0}
  \plotone{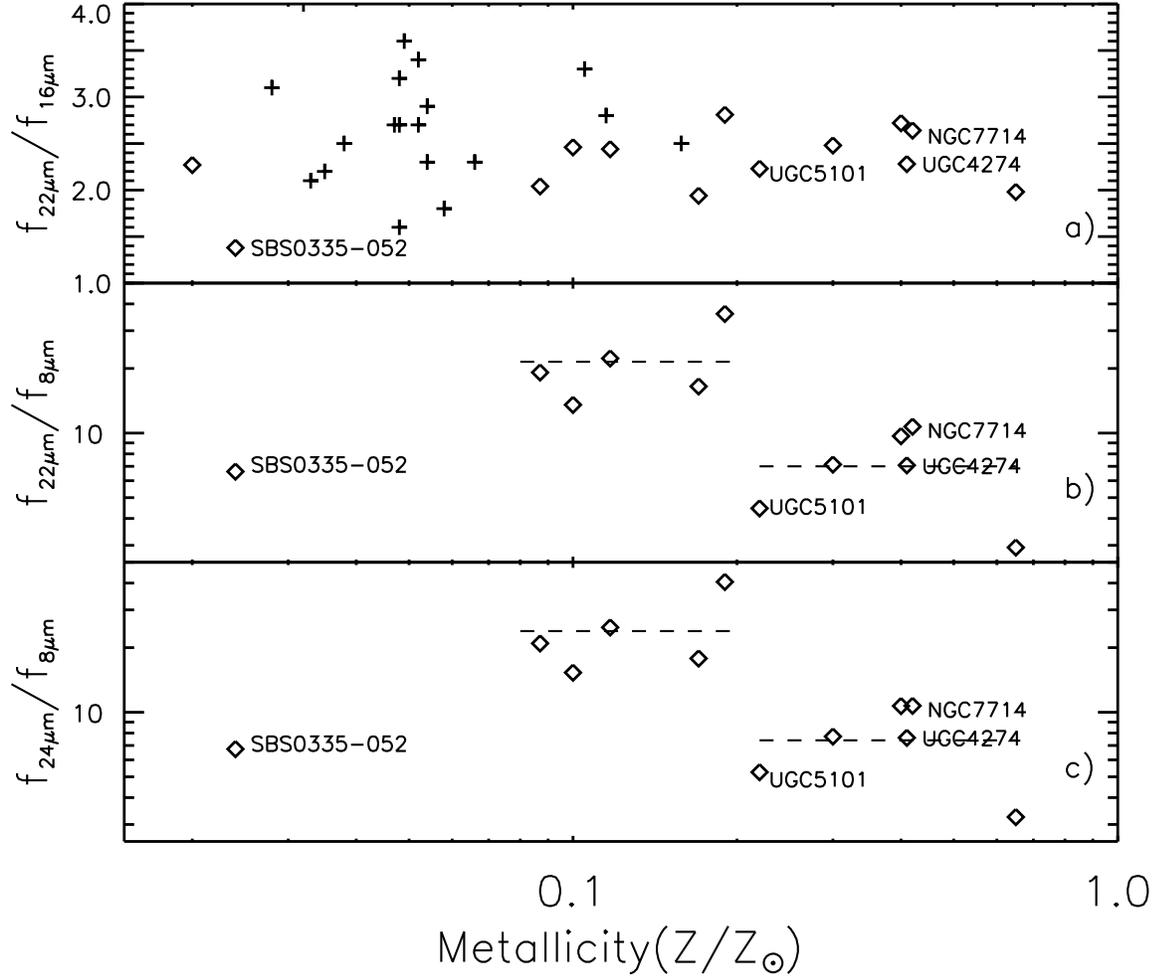}
  \caption{a) This figure shows the ratio of the 22 to the 16\,$\mu$m
    flux density as a function of the metallicity of the sources. For
    sources with an IRS spectrum available, the ratio was calculated
    by convolving the spectrum with the 22 and 16$\mu$m filter curves.
    A starburst galaxy, NGC7714, and a ULIRG, UGC5101, are also
    included for comparison. Diamonds denote sources for which we have
    obtained spectra. The crosses indicate galaxies for which the
    photometric points were obtained using IRS broad band imaging (see
    Section 2). b) The same as a) but for the 22 to 8\,$\mu$m flux
    density as a function of metallicity.  Here we include only the
    galaxies for which we have full {\em IRS} spectra since we measure
    the 8\,$\mu$m flux density from the spectra. SBS0335-052 clearly
    departs from the general trend. We indicate the average flux
    ratios by the dashed lines.c) Same as b) but for 24 over 8\,$\mu$m
    flux density as a function of metallicity.}
  \label{fig:fig5}
\end{figure}

\begin{figure}
  \epsscale{1.0}
  \plotone{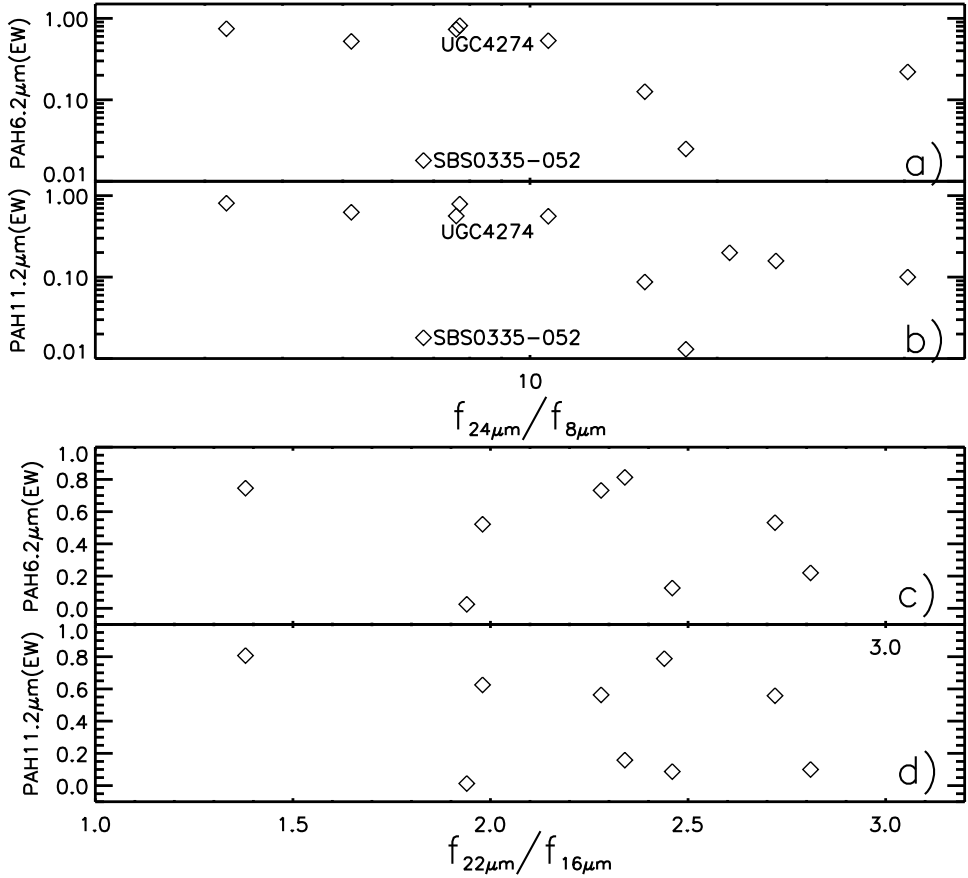}
  \caption{a) This figure presents the ratio of the 6.2\,$\mu$m PAH EW
    of our galaxies as a function of the 24 over 8\,$\mu$m flux
    ratio. We note that with some scatter generally a lower
    $f_{24}$/$f_{8}$ indicates a larger PAH EW. b) Same as a) but for
    the 11.2\,$\mu$m PAH EW.  c) Same as a) but for using the 22 over
    16\,$\mu$m flux density ratio as the variable. There is no clear
    relation between the mid-IR spectral slope as indicated by
    $f_{22}$/$f_{16}$ and the PAH strength. d) Same as c) but for
    11.2\,$\mu$m PAH EW.}
  \label{fig:fig6}
\end{figure}

\begin{figure}
  \epsscale{1.0}
  \plotone{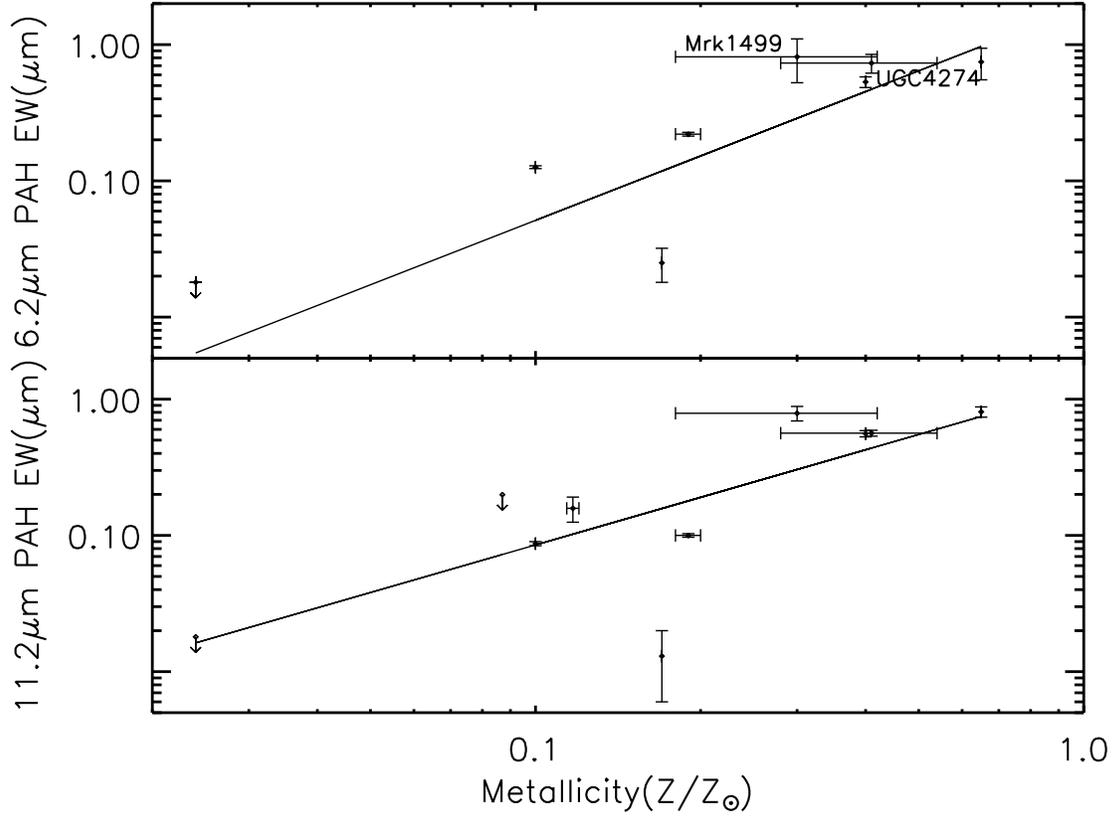}
  \caption{a) This figure shows the PAH EW at 6.2\,$\mu$m vs
    metallicity. Note that the downward arrow indicates the upper
    limit for the PAH emission in SBS0335-052E. In some cases the
    extracted spectra of a target from the two nod positions of the
    slit differ.  These uncertainties are translated into the error
    bars for the EW which, as we can see in Mrk1499, can be fairly
    large (also see text). b) Same as in a) but for the 11.2\,$\mu$m
    PAH feature.}
  \label{fig:fig7}
\end{figure}

\begin{figure}
  \epsscale{1.0}
  \plotone{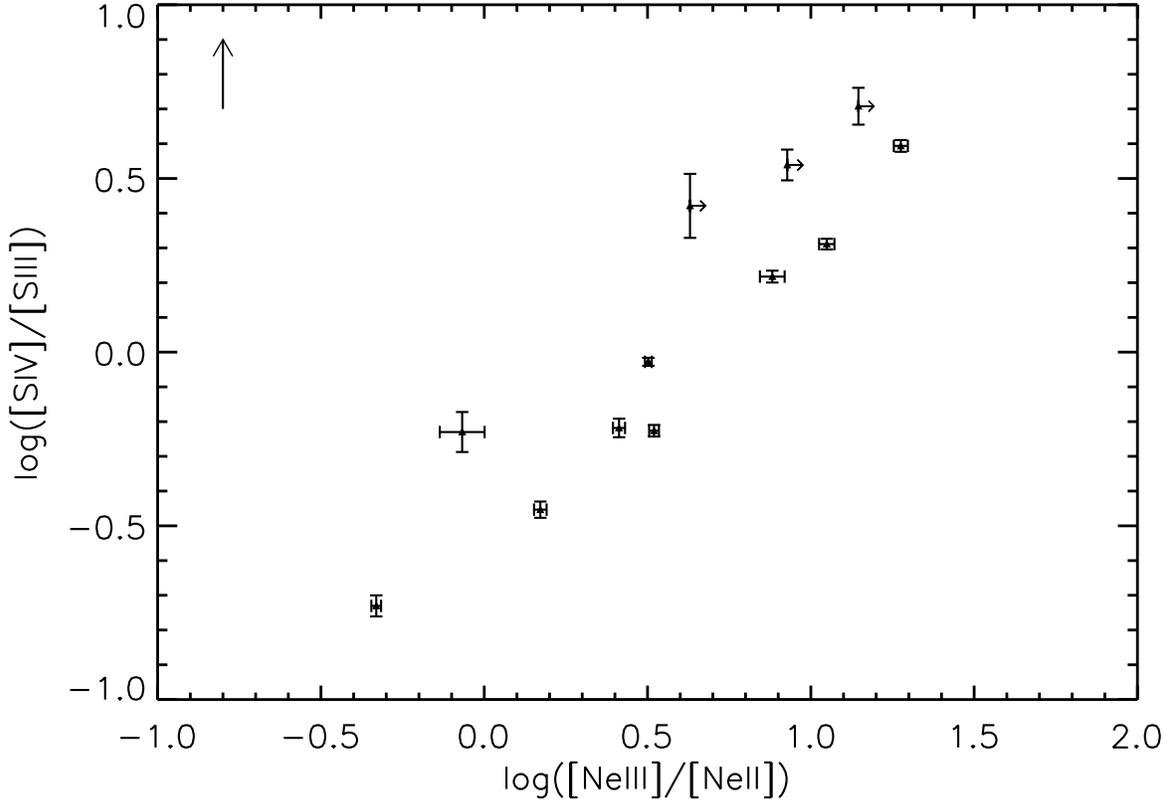}
  \caption{This plot shows the
    log([SIV$\lambda$10.51\,$\mu$m]/[SIII$\lambda$18.71\,$\mu$m]) ratio as
    a function of
    log([NeIII$\lambda$15.55\,$\mu$m]/[NeII$\lambda$12.81\,$\mu$m]) for our
    sample. Both ratios are indicators of the hardness of the
    ionization field.  Since no [NeII] has been detected in the high
    resolution spectrum of SBS0335-052, IZw18 and UM461, the lower
    limits of the [NeIII]/[NeIII] ratios are represented by arrows. No
    correction has been made for extinction. The effect of the
    differential extinction on the ratio of [SIV]/[SIII] is indicated
    by the arrow on the upper left corner for A$_v\sim$13 mag.}
  \label{fig:fig8}
\end{figure}

\begin{figure}
  \epsscale{1.0}
  \plotone{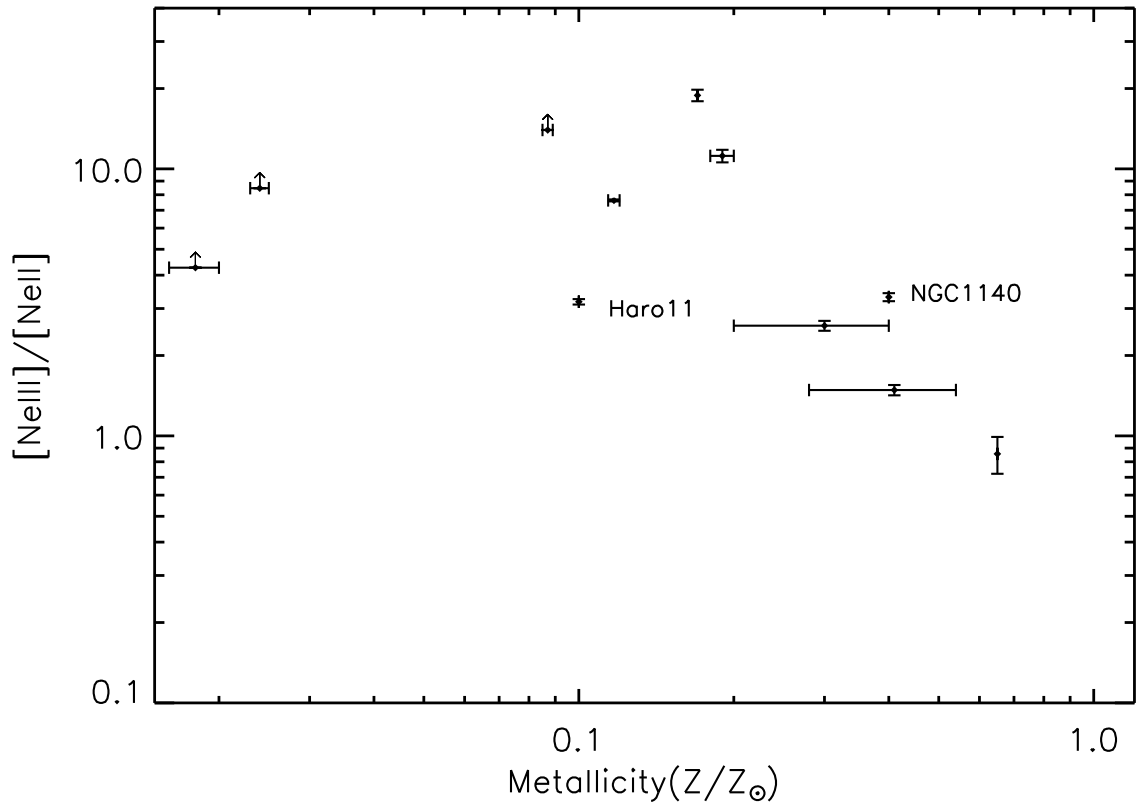}
  \caption{The strength of the [NeIII$\lambda$15.55\,$\mu$m]/[NeII$\lambda$12.81\,$\mu$m] 
    ratio as a function of metallicity for our sample. The fluxes
    of [NeIII] and [NeII] were measured using the high resolution
    spectra of the sources and the metallicity (and its uncertainty)
    were obtained from the literature (see Table 2).  Note that no
    [NeII] has been detected in SBS0335-052E, IZw18 and UM461, so we
    are showing the lower limit of [NeIII]/[NeII] by an up-pointing
    arrow.}
  \label{fig:fig9}
\end{figure}

\begin{figure}
  \epsscale{1.0}
  \plotone{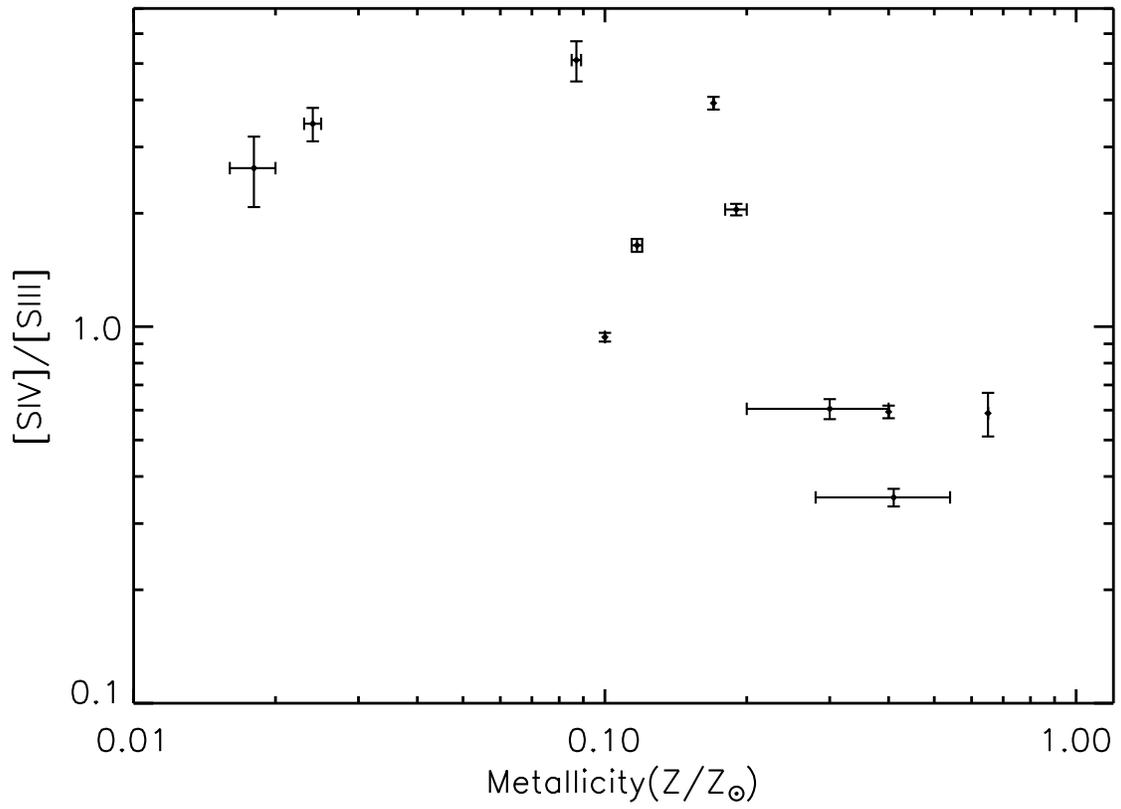}
  \caption{Same as Figure 8 but for
    [SIV$\lambda$10.51\,$\mu$m]/[SIII$\lambda$18.71\,$\mu$m].}
  \label{fig:fig10}
\end{figure}

\begin{figure}
  \epsscale{1.0}
  \plotone{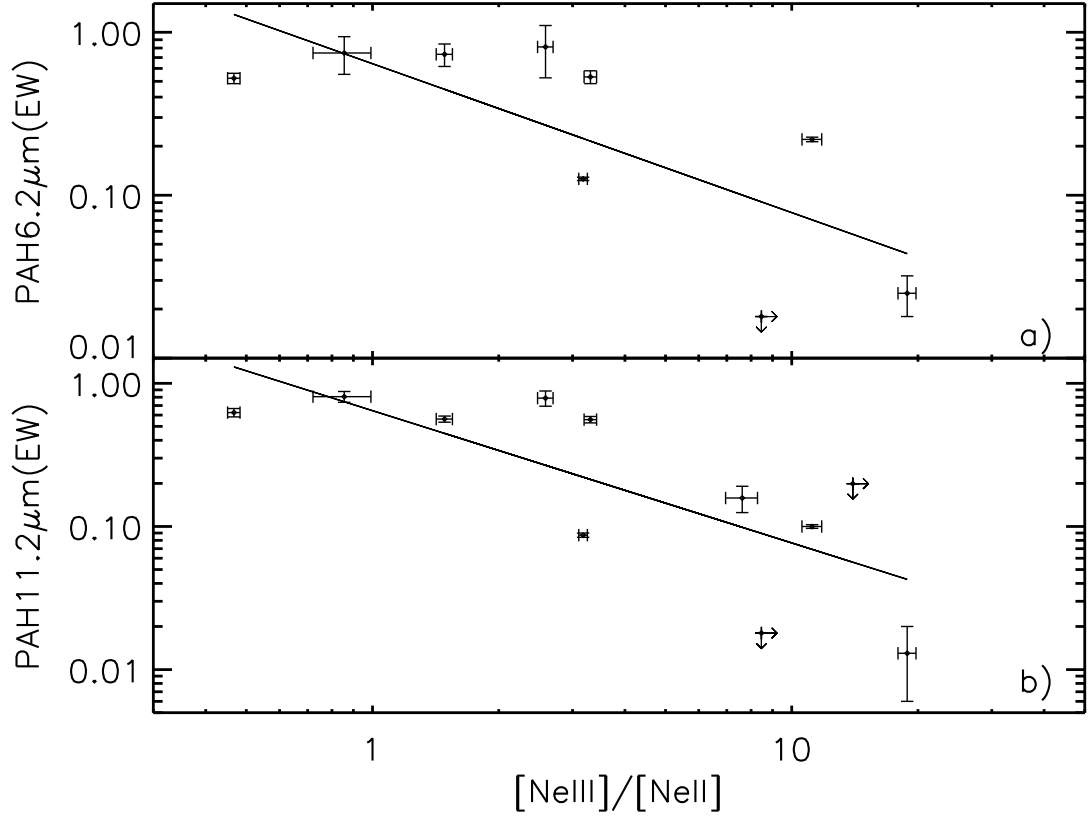}
  \caption{a) The EW of the 6.2\,$\mu$m PAH as a function of the
    [NeIII]/[NeII] ratio. Note that since no PAH emission was detected
    in SBS0335-052E, the upper limit of its EW is indicated by a
    downward arrow. Similarly, the lower limit of the [NeIII]/[NeII]
    ratio for this source is indicated by right-pointing arrow since
    no [NeII] can be seen in the spectrum. b) Same as a) but for the
    11.2\, $\mu$m PAH.}
  \label{fig:fig11}
\end{figure}

\begin{figure}
  \epsscale{1.0}
  \plotone{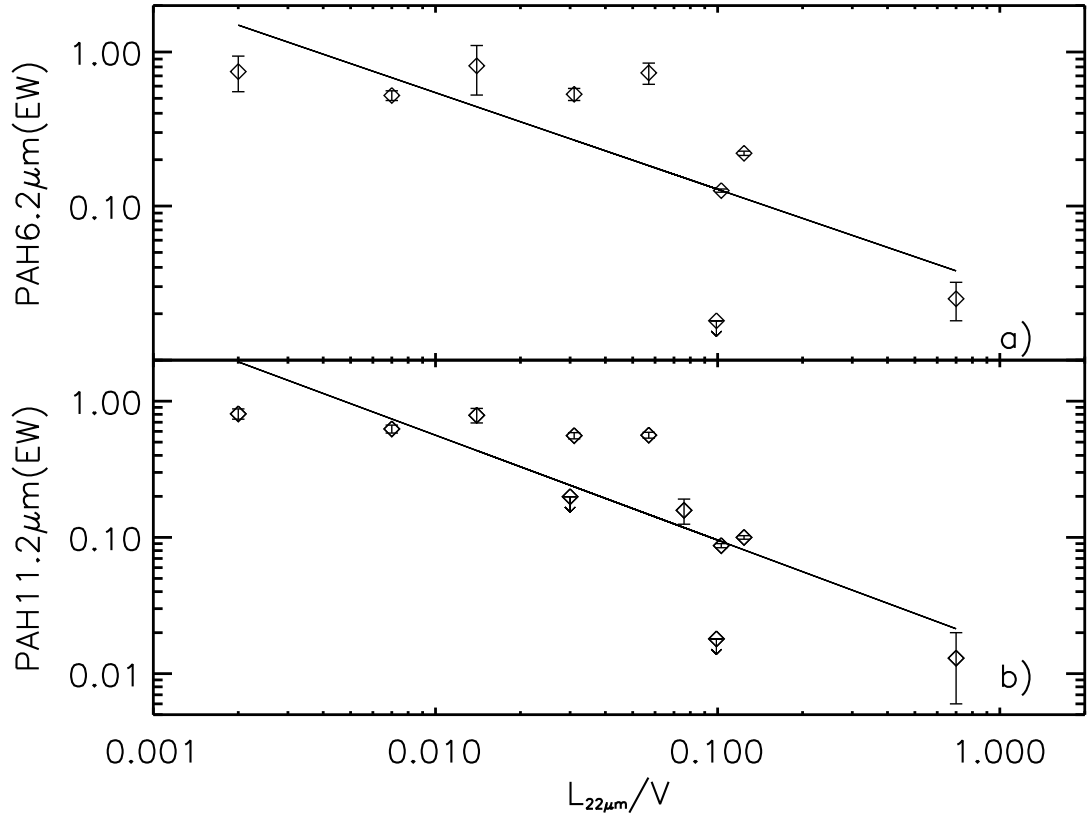}
  \caption{a) The EW of the 6.2\,$\mu$m PAH vs the luminosity density
    indicated by L$_{22\mu m}$/V.  We can see that there is a general trend
    that as the luminosity density of the galaxy increases, the PAH EW
    decreases. b) Same as a) but for the 11.2\,$\mu$m PAH.}
  \label{fig:fig12}
\end{figure}

\begin{figure}
  \epsscale{1.0}
  \plotone{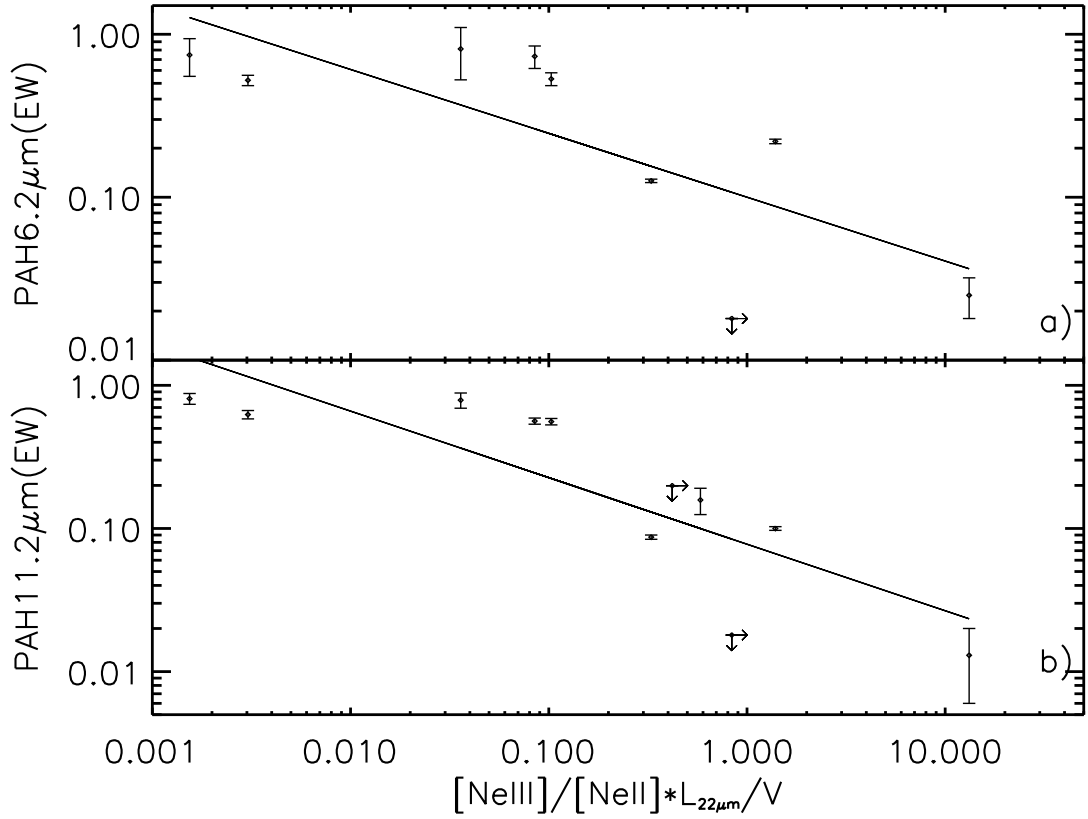}
  \caption{a) The dependence of the 6.2\,$\mu $m PAH EW on the
    ([NeIII]/[NeII])$\times$(L$_{22\mu m}$/V).  It is clear that the
    PAH EW decreases as this quantity increases, indicating that for
    the same physical volume a more luminous starburst with harder
    radiation, destroys the PAHs. b) Same as a) but for the
    11.2\,$\mu$m PAH.}
  \label{fig:fig13}
\end{figure}

\begin{figure}
  \epsscale{1.0}
  \plotone{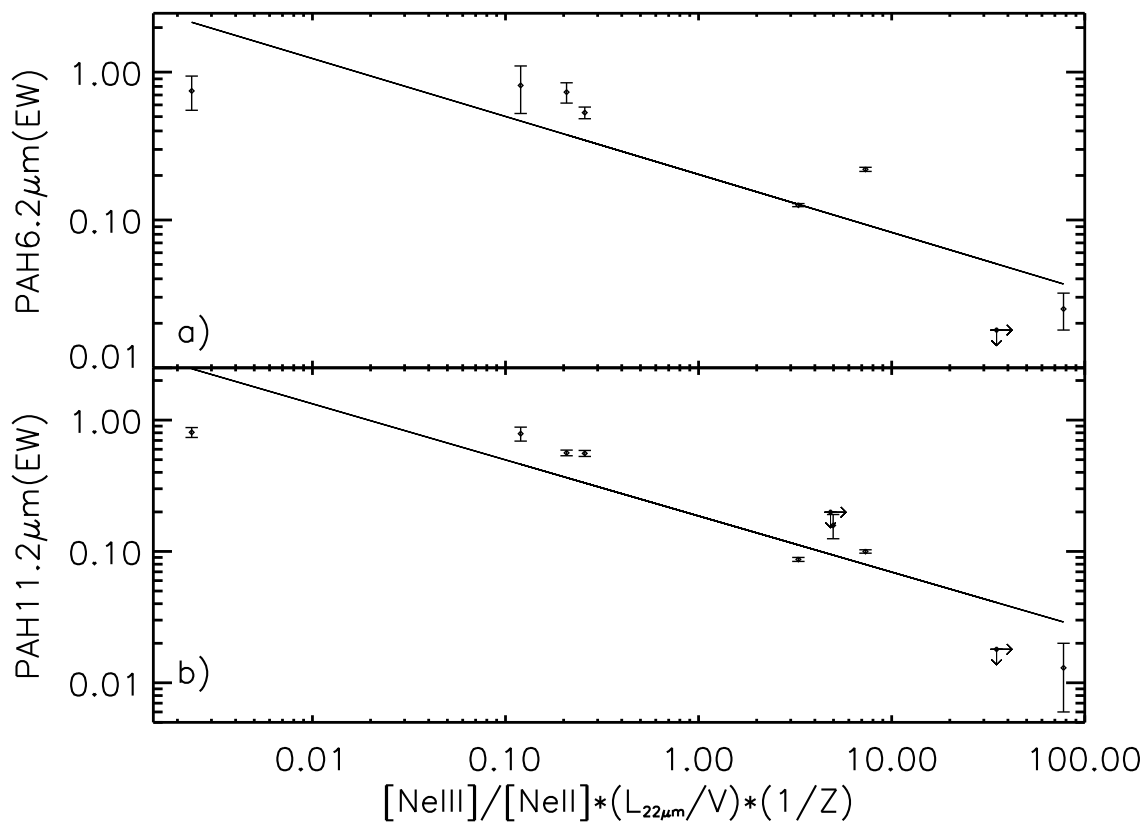}
  \caption{a) The PAH EW at 6.2\,$\mu$m vs the product of the hardness
    of the radiation field and the luminosity density, divided by the
    metallicity of the galaxy. We can see that there is a strong
    anti-correlation where the PAH EW decreases when the factor
    ([NeIII]/[NeII])$\times$(L$_{22\mu m}$/V)$\times$(1/Z)
    increases. b) Same as a) but for the 11.2\,$\mu$m PAH.}
\end{figure}

\clearpage

\begin{deluxetable}{lrllllrrrrrr}
  \tabletypesize{\scriptsize}
  \setlength{\tabcolsep}{0.02in}
  \tablecaption{Properties of Sources\label{tab1}}
  \tablewidth{0pc}
  \tablehead{
    \colhead{Object Name} & \colhead{RA} & \colhead{Dec} & \colhead{AORKEY} &
    \colhead{Observation Date} & \colhead{Redshift} & \multicolumn{6}{c}{On-source Time (sec)}\\
    \colhead{} & \colhead{(J2000)} & \colhead{(J2000)} & \colhead{} & \colhead{} &  \colhead{} & \colhead{SL} &
    \colhead{LL} & \colhead{SH} & \colhead{LH} & \colhead{BPU} & \colhead{RPU}\\
  }
  \startdata
  
Haro11              & 00h36m52.5s  & -33d33m19s &  9007104 & 2004-07-17  &  0.0206 & 168     & 240     & 480     & 240     & \nodata & 98 \\
UM283               & 00h51m49.4s  & +00d33m53s &  8997888 & 2004-07-13  &  0.0155 & \nodata & \nodata & \nodata & \nodata & 28      & 56 \\
UM133               & 01h44m41.3s  & +04d53m26s &  8987392 & 2004-07-16  &  0.0054 & \nodata & \nodata & \nodata & \nodata & 28      & 56 \\
UM382               & 01h58m09.3s  & -00d06m38s &  9004032 & 2005-01-11  &  0.0118 & \nodata & \nodata & 240     & 120     & \nodata & \nodata \\
                    &              &            & 12628224 & 2005-01-14  &         & \nodata & \nodata & \nodata & \nodata & 28      & 56 \\
                    &              &            &  9003776 & 2004-07-16  &         & \nodata & \nodata & \nodata & \nodata & 28      & 56 \\
                    &              &            &  8993792 & 2005-01-11  &         & 56      & 120     & \nodata & \nodata & \nodata & \nodata \\
UM408               & 02h11m23.4s  & +02d20m30s &  8997120 & 2004-08-07  &  0.0120 & \nodata & \nodata & \nodata & \nodata & 28      & 56 \\
NGC1140             & 02h54m33.6s  & -10d01m40s &  4830976 & 2004-01-07  &  0.0050 & 168     & 240     & 480     & 240     & \nodata & 98 \\
SBS0335-052E        & 03h37m44.0s  & -05d02m40s & 11769856 & 2004-09-01  &  0.0135 & \nodata & \nodata & 1440    & 960     & \nodata & 50 \\
                    &              &            &  8986880 & 2004-02-06  &         & 840     & 420     & \nodata & \nodata & \nodata & 50 \\
NGC1569             & 04h30m47.0s  & +64d50m59s &  9001984 & 2004-03-01  & $\sim$ 0& 112     & 120     & 480     & 240     & \nodata & 50 \\
IIZw40              & 05h55m42.6s  & +03d23m32s &  9007616 & 2004-03-01  &  0.0026 & 168     & 240     & 480     & 240     & \nodata & 98 \\
Tol0618-402         & 06h20m02.5s  & -40d18m09s &  4845824 & 2003-12-15  &  0.0350 & \nodata & \nodata & \nodata & \nodata & 28      & 56 \\
SBS0743+591B        & 07h47m46.7s  & +59d00m30s & 12625920 & 2005-03-18  &  0.0211 & \nodata & \nodata & \nodata & \nodata & 28      & 56 \\
                    &              &            & 12622080 & 2005-03-20  &         & \nodata & \nodata & \nodata & \nodata & 28      & 56 \\
SBS0754+570         & 07h58m26.4s  & +56d54m22s & 12630272 & 2005-03-20  &  0.0116 & \nodata & \nodata & \nodata & \nodata & 28      & 56 \\
UGC4274             & 08h13m13.0s  & +45d59m39s & 12076032 & 2004-10-23  &  0.0015 & 112     & 240     & 120     &  56     & \nodata & 50 \\ 
                    &              &            & 12626688 & 2004-11-11  &         & 112     & 240     & 120     &  56     & \nodata & 50 \\ 
SBS0813+582A        & 08h18m04.5s  & +58d05m56s & 12629248 & 2005-03-18  &  0.0268 & \nodata & \nodata & \nodata & \nodata & 28      & 56 \\
HS0822+3542         & 08h25m55.5s  & +35d32m32s & 12630016 & 2004-11-11  &  0.0024 & \nodata & \nodata & \nodata & \nodata & 28      & 56 \\
IZw18               & 09h34m02.0s  & +55d14m28s &  9008640 & 2004-03-27  &  0.0025 & 168     & 240     & 480     & 240     & \nodata & 98 \\
                    &              &            & 12622848 & 2005-04-23  &         & 960     & 480     & \nodata & \nodata & \nodata & 98 \\
SBS0935+495         & 09h38m24.0s  & +49d18m17s & 12624640 & 2004-11-13  &  0.0314 & \nodata & \nodata & \nodata & \nodata & 28      & 56 \\
SBS0940+544         & 09h44m16.7s  & +54d11m33s &  9010432 & 2004-03-26  &  0.0055 & \nodata & \nodata & \nodata & \nodata & 28      & 56 \\
SBS0943+563         & 09h47m13.0s  & +56d06m07s & 12627200 & 2004-11-13  &  0.0253 & \nodata & \nodata & \nodata & \nodata & 28      & 56 \\
SBS1001+555         & 10h04m41.8s  & +55d18m43s & 12624384 & 2005-04-15  &  0.0037 & \nodata & \nodata & \nodata & \nodata & 28      & 56 \\
KUG1013+381         & 10h16m24.5s  & +37d54m46s &  4846336 & 2003-12-15  &  0.0040 & \nodata & \nodata & \nodata & \nodata & 28      & 56 \\
                    &              &            &  8999168 & 2004-04-16  &         & \nodata & \nodata & 120     & 120     & \nodata & \nodata \\
                    &              &            &  9012224 & 2004-04-16  &         & 56      & 240     & \nodata & \nodata & \nodata & \nodata \\
                    &              &            & 12628992 & 2005-04-22  &         & 960     & \nodata & \nodata & \nodata & 50      & \nodata \\
SBS1116+597         & 11h18m47.4s  & +59d26m02s & 12622592 & 2004-11-17  & \nodata & \nodata & \nodata & \nodata & \nodata & 28      & 56 \\
{\rm [RC2]}A1116+51 & 11h19m34.3s  & +51d30m12s &  4846592 & 2003-12-15  &  0.0044 & \nodata & \nodata & \nodata & \nodata & 28      & 56 \\
SBS1119+586         & 11h22m37.8s  & +58d19m43s & 12631040 & 2005-04-15  &  0.0053 & \nodata & \nodata & \nodata & \nodata & 28      & 56 \\
VIIZw403            & 11h27m59.9s  & +78d59m39s &  9005824 & 2004-12-09  & $\sim$ 0& 168     & 240     & 480     & 240     & \nodata & 50 \\
SBS1129+576         & 11h32m02.5s  & +57d22m46s & 12632320 & 2005-01-06  &  0.0052 & \nodata & \nodata & \nodata & \nodata & 28      & 56 \\
SBS1135+598         & 11h37m43.7s  & +59d35m34s & 12624896 & 2004-11-17  &  0.0327 & \nodata & \nodata & \nodata & \nodata & 28      & 56 \\
Mrk1450             & 11h38m35.6s  & +57d52m27s &  9011712 & 2004-12-12  &  0.0032 & 168     & 240     & 480     & 240     & \nodata & 50 \\
SBS1136+607         & 11h39m11.5s  & +60d30m45s & 12622336 & 2004-11-14  &  0.0116 & \nodata & \nodata & \nodata & \nodata & 28      & 56 \\
SBS1137+589         & 11h40m32.0s  & +58d38m32s & 12630528 & 2004-11-17  &  0.0068 & \nodata & \nodata & \nodata & \nodata & 28      & 56 \\
SBS1141+576         & 11h44m16.6s  & +57d24m32s & 12627456 & 2005-04-15  &  0.0310 & \nodata & \nodata & \nodata & \nodata & 28      & 56 \\
UM461               & 11h51m33.3s  & -02d22m22s &  9006336 & 2005-01-03  &  0.0035 & 168     & 240     & 480     & 240     & \nodata & 50 \\
SBS1149+596         & 11h52m34.0s  & +59d22m56s & 12623360 & 2004-11-17  &  0.0112 & \nodata & \nodata & \nodata & \nodata & 28      & 56 \\
SBS1150+599         & 11h53m28.9s  & +59d39m57s & 12621824 & 2004-11-17  &  0.0371 & \nodata & \nodata & \nodata & \nodata & 28      & 56 \\
SBS1159+545         & 12h02m02.4s  & +54d15m50s &  4847104 & 2003-12-15  &  0.0118 & \nodata & \nodata & \nodata & \nodata & 28      & 56 \\
                    &              &            &  9010176 & 2004-04-17  &         & \nodata & \nodata & 120     & 120     & \nodata & \nodata \\
                    &              &            &  9008896 & 2004-04-17  &         & 56      & 240     & \nodata & \nodata & \nodata & \nodata \\
                    &              &            & 12629504 & 2005-04-23  &         & 960     & \nodata & \nodata & \nodata & 50      & \nodata \\
SBS1200+589B        & 12h03m22.6s  & +58d41m36s &  4824064 & 2004-01-06  &  0.0321 & \nodata & 240     & \nodata & \nodata & \nodata & 98 \\
SBS1210+537A        & 12h12m55.9s  & +53d27m38s &  8989952 & 2004-06-06  & \nodata & 168     & 240     & 480     & 240     & \nodata & \nodata \\
SBS1211+564         & 12h13m35.9s  & +56d08m35s & 12623104 & 2005-04-21  &  0.0107 & \nodata & \nodata & \nodata & \nodata & 28      & 56 \\
SBS1212+563         & 12h14m48.5s  & +56d05m19s & 12633088 & 2005-04-21  & \nodata & \nodata & \nodata & \nodata & \nodata & 28      & 56 \\
Tol1214-277         & 12h17m17.1s  & -28d02m33s &  9008128 & 2004-06-28  &  0.0260 & 168     & 240     & 480     & 240     & \nodata & 98 \\
SBS1219+559         & 12h21m29.0s  & +55d38m23s & 12631296 & 2005-04-21  &  0.0308 & \nodata & \nodata & \nodata & \nodata & 28      & 56 \\
SBS1221+545B        & 12h24m23.0s  & +54d14m48s & 12625152 & 2005-04-21  &  0.0187 & \nodata & \nodata & \nodata & \nodata & 28      & 56 \\
HS1222+3741         & 12h24m36.7s  & +37d24m36s & 12630784 & 2005-01-03  &  0.0409 & \nodata & \nodata & \nodata & \nodata & 28      & 56 \\
Tol65               & 12h25m46.9s  & -36d14m01s &  4829696 & 2004-01-07  &  0.0090 & 168     & 240     & 480     & 240     & \nodata & 98 \\
SBS1227+563         & 12h30m07.3s  & +56d05m13s & 12632832 & 2005-04-21  &  0.0153 & \nodata & \nodata & \nodata & \nodata & 28      & 56 \\
{\rm [RC2]}A1228+12 & 12h30m48.5s  & +12d02m42s &  8998656 & 2004-06-27  &  0.0042 & \nodata & \nodata & \nodata & \nodata & 28      & 56 \\
SBS1235+559         & 12h37m36.9s  & +55d41m04s & 12628480 & 2005-04-21  &  0.0293 & \nodata & \nodata & \nodata & \nodata & 28      & 56 \\
UGCA292             & 12h38m40.0s  & +32d46m01s &  4831232 & 2004-01-07  &  0.0010 & 168     & 240     & 480     & 240     & \nodata & 98 \\
                    &              &            & 12076288 & 2005-01-03  &         & \nodata & \nodata & \nodata & \nodata & 28      & 56 \\
Tol1304-353         & 13h07m37.5s  & -35d38m19s &  9006848 & 2004-06-25  &  0.0140 & 168     & 240     & 480     & 240     & \nodata & 50 \\
                    &              &            & 12075520 & 2005-02-10  &         & \nodata & \nodata & \nodata & \nodata & 28      & 56 \\
HS1319+3224         & 13h21m19.9s  & +32d08m23s & 12625664 & 2005-02-07  &  0.0182 & \nodata & \nodata & \nodata & \nodata & 28      & 56 \\
Pox186              & 13h25m48.6s  & -11d37m38s &  9007360 & 2004-07-14  &  0.0039 & 168     & 240     & 480     & 240     & \nodata & 50 \\
                    &              &            & 12625408 & 2005-02-15  &         & \nodata & \nodata & \nodata & \nodata & 28      & 56 \\
SBS1415+437         & 14h17m01.4s  & +43d30m05s &  4844288 & 2004-01-07  &  0.0020 & \nodata & \nodata & \nodata & \nodata & 28      & 56 \\
                    &              &            &  8990464 & 2004-05-15  &         & \nodata & \nodata & 120     & 120     & \nodata & \nodata \\
                    &              &            &  9008384 & 2004-05-13  &         & 56      & 240     & \nodata & \nodata & \nodata & \nodata \\
HS1424+3836         & 14h26m28.1s  & +38d22m59s & 12628736 & 2005-02-13  &  0.0226 & \nodata & \nodata & \nodata & \nodata & 28      & 56 \\
Mrk475              & 14h39m05.4s  & +36d48m22s &  8995840 & 2004-02-07  &  0.0019 & \nodata & \nodata & \nodata & \nodata & 28      & 56 \\
                    &              &            &  8996864 & 2004-06-05  &         & \nodata & \nodata & 120     & 120     & \nodata & \nodata \\
                    &              &            &  8988672 & 2004-06-26  &         & 56      & 240     & \nodata & \nodata & \nodata & \nodata \\
CG0598              & 14h59m20.6s  & +42d16m10s &  8992256 & 2005-03-19  &  0.0575 & 168     & 240     & 480     & 240     & \nodata & 50 \\
CG0752              & 15h31m21.3s  & +47d01m24s &  8991744 & 2005-03-19  &  0.0211 & 168     & 240     & 480     & 240     & \nodata & 50 \\
SBS1533+574B        & 15h34m14.1s  & +57d17m04s & 12627968 & 2005-01-09  &  0.0110 & \nodata & \nodata & \nodata & \nodata & 28      & 56 \\
SBS1538+584         & 15h39m56.9s  & +58d15m33s & 12623616 & 2005-01-08  &  0.0435 & \nodata & \nodata & \nodata & \nodata & 28      & 56 \\
SBS1541+590         & 15h42m55.8s  & +58d55m09s & 12631552 & 2004-12-12  &  0.0450 & \nodata & \nodata & \nodata & \nodata & 28      & 56 \\
Mrk1499             & 16h35m21.1s  & +52d12m53s &  9011456 & 2004-06-05  &  0.0090 & 168     & 240     & 480     & 240     & \nodata & 98 \\
{\rm [RC2]}A2228-00 & 22h30m33.9s  & -00d07m35s &  9006080 & 2004-06-24  &  0.0052 & 168     & 240     & 480     & 240     & \nodata & 50 \\

\enddata

\tablenotetext{a}{The coordinates and redshifts of the objects are cited from The NASA/IPAC Extragalactic Database (NED)(NED is operated 
        by the Jet Propulsion Laboratory, California Institute of Technology, under contract with the National Aeronautics and Space Administration).}
\end{deluxetable}

\clearpage

\begin{deluxetable}{lllllllclll}
  \tabletypesize{\scriptsize}
  \setlength{\tabcolsep}{0.02in}
  \tablecaption{Integrated photometry of Sources\label{tab2}}
   \tablewidth{0pc}
  \tablehead{
    \colhead{Object Name} & \colhead{$f_{8\mu m}$} & \colhead{$f_{16\mu m}$} & \colhead{$f_{22\mu m}$} & \colhead{$f_{24\mu m}$} &
    \colhead{B} & \colhead{K} &  \colhead{$f_{22\mu m}$/$f_{16\mu m}$} &\colhead{Z/Z$_{\odot}$} &\multicolumn{2}{c}{References} \\
    \colhead{} & \colhead{(mJy)} & \colhead{(mJy)} & \colhead{(mJy)} & \colhead{(mJy)} & \colhead{(mag)} & \colhead{(mag)} &
    \colhead{} & \colhead{} &  \colhead{B/K} & \colhead{Z} \\
  }
  \startdata
  \rm   Haro11                       &190.0      &  \nodata  &  851.2\tablenotemark{a}  &1900     & 14.31   & 12.0    & \nodata & 0.1   & 1)~,~2) &19) \\ 
  \rm   UM283                        & \nodata   &  3.0      &    8.2                   & \nodata & 17.38   & \nodata & 2.7     & 0.048 & 9)      &24) \\
  \rm   UM133                        & \nodata   &  2.0      &    6.6                   & \nodata & 15.41   & \nodata & 3.4     & 0.052 & 3)      &25) \\
  \rm   UM382                        & \nodata   &  0.7      &    1.5                   & \nodata & 18.20   & \nodata & 2.2     & 0.035 & 3)      &25) \\
  \rm   UM408                        & \nodata   &  1.3      &    3.5                   & \nodata & 17.46   & \nodata & 2.7     & 0.052 & 3)      &25) \\
  \rm   NGC1140                      & \nodata   &  \nodata  &  294.4                   & \nodata & 12.56   & 10.51   & \nodata & 0.4   & 4)~,~5) &14) \\
  \rm   SBS0335-052E                 & 11.2      &  \nodata  &   78.0                   &  66     & 17.07   & \nodata & \nodata & 0.024 & 10      &17) \\
  \rm   NGC1569                      & \nodata   &  \nodata  & 2067.5\tablenotemark{a}  & \nodata &  9.42   &  7.86   & \nodata & 0.19  & 4)~,~6) &16) \\
  \rm   IIZw40                       &105.0      &  \nodata  &  796.1\tablenotemark{a}  &1500     & 11.87   & 12.35   & \nodata & 0.17  & 3)~,~5) &15) \\
  \rm   SBS0743+591B                 & \nodata   &  2.3      &    4.7                   & \nodata & \nodata & \nodata & 2.1     &\nodata \\
  \rm   SBS0754+570                  & \nodata   &  0.8      &    2.4                   & \nodata & \nodata & \nodata & 2.9     &\nodata \\
  \rm   UGC4274                      & \nodata   &  \nodata  &  130.2                   & \nodata & 12.07   &  9.13   & \nodata & 0.41  & 4)~,~6) &18)\tablenotemark{g} \\
  \rm   HS0822+3542                  &  0.2      &  1.3      &    4.1                   &   2.7   & 17.92   & \nodata & 3.1     & 0.028 & 26)     &26) \\
  \rm   IZw18                        &  0.7      &  \nodata  &    4.1                   &   5.5   & 16.05   & 15.92   & \nodata & 0.018 & 3)~,~11)&27) \\
  \rm   SBS0935+495                  & \nodata   &  1.2      &    2.9                   & \nodata & \nodata & \nodata & 2.3     &\nodata \\
  \rm   SBS0940+544                  & \nodata   &  1.1      &    2.3                   & \nodata & 17.18   & \nodata & 2.1     & 0.033 & 3)      &21) \\
  \rm   SBS0943+563                  & \nodata   &  1.4      &    3.8                   & \nodata & \nodata & \nodata & 2.8     &\nodata \\
  \rm   KUG1013+381                  & \nodata   &  6.4      &   17.1                   & \nodata & 16.04   & \nodata & 2.7     & 0.047 & 28)     &28) \\
  \rm   SBS1116+597                  & \nodata   &  1.0      &    1.1                   & \nodata & \nodata & 13.28   & 1.2     &\nodata& 5) \\
  \rm   [RC2]A1116+51                & \nodata   &  0.5      &    1.7                   & \nodata & 17.46   & \nodata & 3.6     & 0.049 & 12)     &29) \\
  \rm   SBS1119+586                  & \nodata   &  0.6      &    1.5                   & \nodata & \nodata & \nodata & 2.6     &\nodata \\
  \rm   VIIZw403                     &  2.5      &  \nodata  &    7.7                   &  28     & 14.11   & 12.7    & \nodata & 0.060 & 3)~,~7) &21) \\
  \rm   SBS1129+576                  & \nodata   &  0.3      &    1.4                   & \nodata & \nodata & \nodata & 4.0     & 0.032 &         &30)\\
  \rm   SBS1135+598                  & \nodata   &  0.2      &    0.8                   & \nodata & \nodata & \nodata & 3.7     &\nodata \\
  \rm   Mrk1450                      &  2.4      &  \nodata  &   52.1                   &  48     & 15.75   & \nodata & \nodata & 0.117 & 3)      &21) \\
  \rm   SBS1136+607                  & \nodata   &  0.5      &    1.8                   & \nodata & \nodata & \nodata & 3.8     &\nodata \\
  \rm   SBS1137+589                  & \nodata   &  0.5      &    1.0                   & \nodata & \nodata & \nodata & 2.2     &\nodata \\
  \rm   SBS1141+576                  & \nodata   &  1.1      &    2.9                   & \nodata & \nodata & \nodata & 2.5     &\nodata \\
  \rm   UM461                        &  1.6      &  \nodata  &   29.2                   &  30     & \nodata & \nodata & \nodata & 0.087 &         &22) \\
  \rm   SBS1149+596                  & \nodata   &  1.5      &    4.4\tablenotemark{b}  & \nodata & \nodata & \nodata & 3.1     &\nodata \\
  \rm   SBS1150+599                  & \nodata   &  0.3      &    0.6                   & \nodata & \nodata & 12.67   & 1.8     &       & 5) \\
  \rm   SBS1159+545                  & \nodata   &  2.6      &    6.4                   & \nodata & \nodata & \nodata & 2.5     & 0.038 &         &21) \\
  \rm   SBS1212+563                  & \nodata   &  1.0      &    1.2                   & \nodata & \nodata & \nodata & 1.2     &\nodata \\
  \rm   Tol1214-277                  &  0.2      &  \nodata  &    3.5                   &   5.5   & \nodata & \nodata & \nodata & 0.045 &         &24)\\
  \rm   SBS1219+559                  & \nodata   &  1.0      &    2.1                   & \nodata & \nodata & \nodata & 2.1     &\nodata \\
  \rm   SBS1221+545                  & \nodata   &  0.6      &    1.9                   & \nodata & \nodata & \nodata & 3.0     &\nodata \\
  \rm   HS1222+3741                  & \nodata   &  3.0      &    6.9                   & \nodata & 18.40   & \nodata & 2.3     & 0.054 & 13)     &32)\\
  \rm   Tol65                        &  0.9      &  \nodata  &   15.7                   &  15     & 17.26   & \nodata & \nodata & 0.031 & 3)      &24) \\
  \rm   SBS1227+563                  & \nodata   &  1.0      &    1.7                   & \nodata & \nodata & \nodata & 1.7     &\nodata \\
  \rm   [RC2]A1228+12                & \nodata   &  1.2      &    3.4                   & \nodata & 17.96   & \nodata & 2.9     & 0.054 & 8)      &29) \\
  \rm   SBS1235+559                  & \nodata   &  0.3      &    1.0                   & \nodata & \nodata & \nodata & 2.9     &\nodata \\
  \rm   Tol1304-353                  & \nodata   &  6.0      &   10.8                   & \nodata & \nodata & \nodata & 1.8     & 0.058 &         &33) \\
  \rm   HS1319+3224-                 & \nodata   &  1.3      &    2.1                   & \nodata & 19.0    & \nodata & 1.6     & 0.048 & 13)     &32) \\
  \rm   Pox186                       & \nodata   &  5.8      &   13.1                   & \nodata & 17.0    & \nodata & 2.3     & 0.066 & 14)     &31)\\
  \rm   SBS1415+437                  & \nodata   &  6.0      &   19.6                   & \nodata & 15.43   & \nodata & 3.2     & 0.048 & 3)      &21) \\
  \rm   HS1424+3836-                 & \nodata   &  0.5      &    1.5                   & \nodata & 18.68   & \nodata & 2.8     & 0.115 & 13)     &32)\\
  \rm   Mrk475                       & \nodata   &  3.2      &   10.8                   & \nodata & 16.20   & \nodata & 3.3     & 0.105 & 3)      &21) \\
  \rm   CG0598                       & \nodata   &  \nodata  &   23.0                   & \nodata & \nodata & 13.17   & \nodata & 0.65  & 5)      &23) \\
  \rm   CG0752                       & \nodata   &  \nodata  &  138.9                   & \nodata & \nodata & 11.73   & \nodata &\nodata& 5) \\
  \rm   SBS1533+574B                 & \nodata   & 21.2      &   53.4                   & \nodata & 16.02   & \nodata & 2.5     & 0.158 & 3)      &21) \\
  \rm   SBS1538+584                  & \nodata   &  1.4      &    3.9                   & \nodata & \nodata & \nodata & 2.8     &\nodata \\
  \rm   SBS1541+590                  & \nodata   &  1.3      &    3.3                   & \nodata & \nodata & \nodata & 2.5     &\nodata\\
  \rm   Mrk1499                      & \nodata   &  \nodata  &   29.5                   & \nodata & 16.00   & \nodata & \nodata & 0.3   & 4)      &20)~\tablenotemark{g} \\
  
  \enddata
  
  \tablenotetext{a}{The 22$\mu$m peak-up is saturated.}
  \tablenotetext{b}{The error in this value is higher $\sim$10\% in the blue  
    and $\sim$15\% in the red due to the presence of a bright star.}
  \tablenotetext{c}{\nodata indicates that no data are available.}
  \tablenotetext{d}{The {\em IRAC} 8\,$\mu$m and {\em MIPS} 24\,$\mu$m flux 
    densities are quoted from \citet{Engelbracht2005}.}
  \tablenotetext{e}{References for B and K magnitude: 
    1) \citet{Lauberts1989}, 2) \citet{Spinoglio1995}, 3) \citet{GildePaz2003},
    4) \citet{DeVaucoulers1991}, 5) 2Mass Extended Objects. Final Release (2003), 
    6)\citet{Jarrett2003}, 7) \citet{Tully1981}, 8) \citet{Young1998},
    9) \citet{Vitores1996}, 10) \citet{Papaderos1998}, 11) \citet{Thuan1983},
    12) \citet{Arp1975}, 13) \citet{Vennik2000},
    26) \citet{Kniazev2000}, 28) \citet{Kniazev1998}}
  \tablenotetext{f}{References for metallicity: The metallicity we list 
    here is converted from the oxygen abundance using 12+log[O/H]=8.91 for solar
    metallicity \citep{Kunth2000}. 14) \citet{Calzetti1997}, 15) \citet{Cervino1994}, 
    16) \citet{Kobulnicky1997}, 17) \citet{Izotov1997b}, 18) \citet{Ho1997}, 
    19) \citet{Bergvall2000}, 20) \citet{Petrosian2002}, 21) \citet{Izotov1999}, 
    22) \citet{Kniazev2004}, 23) \citet{Peimbert1992}, 24) \citet{Gallego1997}, 
    25) \citet{Masegosa1994}, 26) \citet{Kniazev2000}, 27) \citet{Skillman1993}, 
    28) \citet{Kniazev1998}, 29) \citet{Kinman1981}, 30) \citet{Guseva2003a}, 
    31) \citet{Guseva2003b}, 32) \citet{Popescu2000}, 33) \citet{Stasinska1996}}
  \tablenotetext{g}{The oxygen abundance is calculated from [NII]/H$\alpha$ measurement 
    using the method proposed by \citet{Denicolo2002}} 
\end{deluxetable}

\clearpage

\begin{deluxetable}{lcccc}
  \tabletypesize{\scriptsize}
  \setlength{\tabcolsep}{0.02in}
  \tablecaption{Synthetic flux density of the sources\label{tab3}}
  \tablewidth{0pc}
  \tablehead{
    \colhead{Object Name} & \colhead{{\em IRAC} 8\,$\mu$m} & \colhead{{\em IRS} 16\,$\mu$m (BPU)} & \colhead{{\em IRS} 22\,$\mu$m (RPU)} &
    \colhead{{\em MIPS} 24\,$\mu$m}\\
    \colhead{} & \colhead{(mJy)} & \colhead{(mJy)} & \colhead{(mJy)} & \colhead{(mJy)} \\
  }
  \startdata

  \rm IZw18       &   \nodata     &    2.6$\pm0.1$ &    5.9$\pm0.3$ &    6.5$\pm0.1$ \\  
  \rm SBS0335-052 &  12.3$\pm0.2$ &   58.1$\pm0.3$ &   79.8$\pm0.3$ &   81.5$\pm0.2$ \\
  \rm UM461       &   1.8$\pm0.1$ &   16.9$\pm0.3$ &   34.5$\pm0.2$ &   37.7$\pm0.2$ \\
  \rm Haro11      & 164.2$\pm0.1$ &  932.6$\pm0.3$ & 2307.0$\pm4.5$ & 2640.4$\pm6.5$ \\ 
  \rm Mrk1450     &   2.4$\pm0.2$ &   22.0$\pm0.1$ &   53.5$\pm0.2$ &   59.7$\pm0.1$ \\ 
  \rm IIZw40      &  90.1$\pm0.6$ &  763.7$\pm2.4$ & 1481.9$\pm6.9$ & 1624.0$\pm11.5$ \\
  \rm NGC1569     &  72.2$\pm0.2$ &  921.8$\pm0.6$ & 2596.0$\pm1.0$ & 2991.0$\pm4.6$ \\
  \rm Mrk1499     &   4.2$\pm0.1$ &   12.4$\pm0.1$ &   30.4$\pm0.1$ &   33.8$\pm0.1$ \\
  \rm NGC1140     &  28.7$\pm0.3$ &  103.5$\pm0.1$ &  279.5$\pm0.3$ &  316.6$\pm0.2$ \\ 
  \rm UGC4274     &  17.0$\pm0.2$ &   53.1$\pm0.2$ &  120.5$\pm0.7$ &  132.5$\pm0.5$ \\  
  \rm CG0598      &   7.2$\pm0.1$ &   11.2$\pm0.1$ &   24.8$\pm0.1$ &   28.3$\pm0.1$ \\
  \rm CG0752      &  32.9$\pm0.6$ &   63.6$\pm0.2$ &  148.6$\pm0.1$ &  169.9$\pm0.1$ \\

  \enddata
\end{deluxetable}

\begin{deluxetable}{lcccccccc}
  \tabletypesize{\scriptsize}
  \setlength{\tabcolsep}{0.02in}
  \tablecaption{Spectroscopic Data of the Sources\label{tab4}}
  \tablewidth{0pc}
  \tablehead{
    \colhead{Object Name} & \multicolumn{4}{c}{PAH EW ($\mu$m)} & \multicolumn{4}{c}{Flux\tablenotemark{a} ($\times10^{-16}$W m$^{-2}$)}\\
    \colhead{} & \colhead{6.2\,$\mu$m} & \colhead{7.7\,$\mu$m} & \colhead{8.6\,$\mu$m} & 
    \colhead{11.2\,$\mu$m} & \colhead{[SIV](10.51\,$\mu$m)} & \colhead{[NeII](12.81\,$\mu$m)} & 
    \colhead{[NeIII](15.55\,$\mu$m)} & \colhead{[SIII](18.71\,$\mu$m)} \\
  }
  \startdata

  \rm IZw18       &  \nodata           & \nodata           & \nodata           &   \nodata          &  0.06$\pm0.01$    &$<$0.011            &  0.05$\pm0.01$    &   0.021$\pm0.001$ \\  
  \rm SBS0335-052 &$<$0.018            & \nodata           & \nodata           &$<$0.018            &  0.16$\pm0.01$    &   $<$0.016         &  0.135$\pm0.004$  &   0.045$\pm0.004$ \\
  \rm UM461       &  \nodata           & \nodata           & \nodata           & $<$0.199           &  0.47$\pm0.03$    &$<$0.021            &  0.29$\pm0.01$    &   0.09$\pm0.01$ \\
  \rm Haro11      &   0.126$\pm0.003$  &  0.241$\pm0.006$  &  0.055$\pm0.003$  &   0.087$\pm0.003$  &  4.6$\pm0.1$      &   3.2$\pm0.1$      & 10.1$\pm0.1$      &   4.81$\pm0.02$ \\ 
  \rm Mrk1450     &  \nodata           & \nodata           & \nodata           &   0.158$\pm0.033$  &  0.75$\pm0.02$    &   0.13$\pm0.01$    &  0.96$\pm0.02$    &   0.46$\pm0.01$ \\ 
  \rm IIZw40      &   0.025$\pm0.007$  &  0.069$\pm0.005$  &  0.009$\pm0.003$  &   0.013$\pm0.007$  & 18.6$\pm0.4$      &   0.60$\pm0.03$    & 11.4$\pm0.3$      &   4.7$\pm0.2$ \\
  \rm NGC1569     &   0.220$\pm0.007$  &  0.402$\pm0.012$  &  0.045$\pm0.004$  &   0.100$\pm0.003$  & 14.8$\pm0.4$      &   1.55$\pm0.08$    & 17.4$\pm0.3$      &   7.2$\pm0.2$ \\
  \rm Mrk1499     &   0.831$\pm0.288$  &  0.997$\pm0.189$  &  0.345$\pm0.049$  &   0.788$\pm0.096$  &  0.18$\pm0.01$    &   0.19$\pm0.01$    &  0.50$\pm0.02$    &   0.30$\pm0.01$ \\
  \rm NGC1140     &   0.532$\pm0.048$  &  0.579$\pm0.025$  &  0.142$\pm0.008$  &   0.558$\pm0.029$  &  1.30$\pm0.04$    &   1.17$\pm0.02$    &  3.8$\pm0.1$      &   2.17$\pm0.04$ \\
  \rm UGC4274     &   0.732$\pm0.115$  &  0.671$\pm0.040$  &  0.191$\pm0.010$  &   0.563$\pm0.028$  &  0.46$\pm0.02$    &   0.88$\pm0.03$    &  1.31$\pm0.04$    &   1.30$\pm0.03$ \\ 
  \rm CG0598      &   0.746$\pm0.194$  &  0.954$\pm0.075$  &  0.331$\pm0.036$  &   0.807$\pm0.070$  &  0.13$\pm0.02$    &   0.206$\pm0.004$  &  0.18$\pm0.03$    &   0.23$\pm0.01$ \\
  \rm CG0752      &   0.522$\pm0.038$  &  0.645$\pm0.024$  &  0.171$\pm0.008$  &   0.625$\pm0.042$  &  0.15$\pm0.01$    &   1.23$\pm0.03$    &  0.58$\pm0.01$    &   0.82$\pm0.02$ \\

  \enddata
  \tablenotetext{a}{The integrated fluxes are measured from the high resolution spectra of the targets.
    Background emission hasn't been subtracted and no scaling between the two modules, SH and LH has 
    been applied.}
  \tablenotetext{b}{\nodata indicates no available PAH EW measurement. It is mostly due to the low SNR 
    of the spectrum. In this case, the determination of the continuum is highly uncertain and will significantly 
    change the PAH EW. A much deeper observation with more exposure time has been submitted and more analysis 
    will follow when data come in. For SBS0335-052, the SNR is high enough, but no PAH features can be identified 
    in its mid-IR spectrum. We do not have a template to derive the upper limit of PAH EW at 7.7\,$\mu$m and 8.6\,$\mu$m.}
  
\end{deluxetable}

\clearpage

\begin{deluxetable}{lrrrrr}
  \tabletypesize{\scriptsize}
  \setlength{\tabcolsep}{0.02in}
  \tablecaption{Galaxy luminosity and volume\label{tab5}}
  \tablewidth{0pc}
  \tablehead{
    \colhead{Object} & \colhead{$f_{22\mu m}$\tablenotemark{a}} & \colhead{Radius\tablenotemark{b}} & \colhead{$L_{22\mu m}$\tablenotemark{c}} &
    \colhead{Volume} & \colhead{$L_{22\mu m}$/Volume} \\
    \colhead{} & \colhead{(Jy)} & \colhead{(arcsec)} & \colhead{($\times10^{8}L_{\odot}$)} & \colhead{($\times10^{9}pc^{3}$)} & \colhead{(L$_{\odot}/pc^{3}$)} \\
  }    
  \startdata

  \rm IZw18       & 0.006 &  7.2 &  0.01 &   0.2  & 0.007 \\
  \rm SBS0335-052 & 0.080 &  3.8 &  3.8  &   3.9  & 0.099 \\
  \rm UM461       & 0.030 &  6.8 &  0.1  &   0.4  & 0.030 \\  
  \rm Haro11      & 2.307 &  9.9 &249.7  & 241.4  & 0.103 \\
  \rm Mrk1450     & 0.053 &  5.9 &  0.1  &   0.2  & 0.076 \\
  \rm IIZw40      & 1.482 &  9.5 &  2.1  &   0.3  & 0.700 \\
  \rm NGC1569     & 2.596 & 55.4 &  0.2  &   0.1  & 0.124 \\
  \rm Mrk1499     & 0.030 &  6.1 &  0.6  &   4.6  & 0.014 \\
  \rm NGC1140     & 0.279 & 28.0 &  1.8  &   5.8  & 0.031 \\
  \rm UGC4274     & 0.120 & 30.5 &  0.1  &   0.1  & 0.057 \\
  \rm CG0598      & 0.025 &  5.9 & 20.8  &1185.6  & 0.002 \\
  \rm CG0752      & 0.149 &  9.9 & 16.9  & 258.3  & 0.007 \\

  \enddata
  \tablenotetext{a}{The 22\,$\mu$m flux density are calculated using the ``synthetic'' method.}
  \tablenotetext{b}{The radius of the galaxy is measured using the optical image retrieved from 
    the digital sky server(DSS) for point sources. For the four extended sources mentioned earlier, 
    we used their mid-IR size measured from the 22\,$\mu$m images.}
  \tablenotetext{c}{The 22\,$\mu$m luminosity in the red peak-up filter(18.5-26.0\,$\mu$m).}
  
\end{deluxetable}

\end{document}